\documentclass[prd,twocolumn,floatfix,amsmath,nofootinbib,amssymb,floatfix]{revtex4}
\usepackage{amsmath}
\usepackage{graphicx}
\usepackage{bm}
\usepackage{epsf}

\usepackage{amsmath,amssymb}
\usepackage{graphicx,bm}
\usepackage{slashed}
\usepackage{epstopdf}
\usepackage{ulem} 
\usepackage[usenames]{color}
\usepackage{float}
\usepackage[colorlinks,
            citecolor=blue,
            anchorcolor=red,
            menucolor=red,
            linkcolor=red,
            filecolor=red,
            runcolor=red,
            urlcolor=blue,
            frenchlinks=red]{hyperref}
\usepackage{subfigure}
\usepackage{rotating}
\usepackage{multirow}
\usepackage{dcolumn}
\usepackage{overpic}
\usepackage{booktabs}
\usepackage{makecell}

\usepackage{xcolor}

\begin{document}

\title{Properties of H particle-admixed compact star}

\author{Xuhao Wu$^1$}
\author{Liming Wang $^{1,3}$}\email{lmwang@ysu.edu.cn}
\author{Hong-Tao An$^{2,3}$}\email{anht@tsinghua.edu.cn}
\author{Min Ju$^4$}\email{jumin@upc.edu.cn}
\author{Hong Shen$^5$}\email{shennankai@gmail.com}
\affiliation{
$^1$Key Laboratory for Microstructural Material Physics of Hebei Province, School of Science, Yanshan University, Qinhuangdao 066004, China\\
$^2$ Department of Physics, Tsinghua University, Beijing 100084, China \\
$^3$Lanzhou Center for Theoretical Physics, Key Laboratory of Theoretical Physics of Gansu Province,
and Frontiers Science Center for Rare Isotopes, Lanzhou University, Lanzhou 730000, China\\
$^4$School of Science, China University of Petroleum (East China), Qingdao 266580, China\\
$^5$School of Physics, Nankai University, Tianjin, 300071, China
}

\begin{abstract}
We explore the potential manifestation of a hexaquark, the H particle, as a constituent within neutron stars.
The H particle, with a quark content of $uuddss$, is constructed within the Chromomagnetic Interaction (CMI) framework.
Specifically, we contemplate the flavor-singlet state H with $J^P=0^+$.
Our computations indicate that the three-flavor hexaquark state, the H particle, possesses a lower mass of $2212.7~\rm{MeV}$ in comparison to the $d^*(2380)$,
implying greater stability than the two-flavor $d^*(2380)$.
The analysis involving the H particle is carried out using
the relativistic mean-field (RMF) model.
We investigate the influence of H particle couplings,
a key factor in determining the system stability,
and focus on the potential existence of H particle within neutron stars.
We find that H particle could potentially endure as a stable constituent within neutron stars,
and lead to a reduction of the maximum mass.

\end{abstract}


\maketitle

\section {Introduction}
In 1977, Jaffe~\cite{Jaffe1977} proposed the H-dibaryon, a spin-0, flavor-singlet, parity-even $uuddss$ state with electric charge $Q=0$, baryon number $B=2$ and strangeness number $S=-2$.
Within the MIT bag model, this state was predicted to be a stable six-quark configuration and represents the earliest theoretical realization of a bound hexaquark.
In the broader context of six-quark systems, several distinct $uuddss$ states have been discussed, 
including Jaffe's original H-dibaryon and the deeply bound sexaquark~\cite{Farrar2018} proposed by Farrar as a dark matter candidate.
These states, though sharing the same quark content and flavor quantum numbers, differ significantly in their binding mechanism and mass, and therefore should not be regarded as equivalent~\cite{Shahrbaf2022}.
The H particle considered in the present work corresponds to a flavor-singlet, spin-0 $uuddss$ hexaquark with a mass of 2212.7 MeV, lies below the $\Lambda\Lambda$ threshold ($\approx 2231$ MeV).
Such a mass implies that the H particle would be stable in hadronic matter and could therefore emerge as a new degree of freedom at the high baryon densities reached in neutron-star (NS) interiors.

Traditionally, the core of NS has been modeled as a uniform fluid
composed of neutron-rich nuclear matter in $\beta$ equilibrium.
However, at the extremely high densities in NS core, it is expected that new degrees of freedom  will emerge beyond nucleons~\cite{Weber2007}.
The candidates can include  hyperons~\cite{Panda2004,Stone2007,Katayama2015,Li2018,Huang2022,Drago2014,Sahoo2018}, $\Delta-$isobars~\cite{Drago2014,Zhu2016,Sahoo2018,Li2020,Raduta2021,Mantziris2020},
di-baryonic matter~\cite{Faessler1997,Faessler1997b,Faessler1998,Glendenning1998,Mantziris2020} and
deconfined quark matter~\cite{Glendenning2001,Heiselberg2000,Weber2005,Nakazato2008,Benic2014,Wu2019,Ju2021}.
The presence of these exotic degrees of freedom in NSs are generally in competition with other constitutes,
affecting all particles' fraction,
softening the equation of state (EOS),
and reducing the maximum mass of the NS.
The existence of the dibaryon in NSs is possible under high chemical potentials,
and its bosonic nature sets it apart from the normal fermionic constituents such as nucleons and leptons.
Extensive studies have been conducted on the hexaquark $d^*(2380)$ ~\cite{Vidana2018,Bashkanov2019,Kim2020,Celi2024},
supported by experimental evidence~\cite{Adlarson2014}.
$d^*(2380)$ shares the same composition of $u$, $d$ quarks and does not involve any strangeness degrees of freedom.
Despite its large mass, $d^*(2380)$ still can exist in the interior of NSs~\cite{Mantziris2020}.
At high densities, three flavor quark matter may be more stable than two flavor quark matter~\cite{Witten1984,Liu2023}.
The H particle, possessing three-flavor symmetry, is expected to be more stable than $d^*(2380)$ with lower mass and energy per baryon,
making it capable of stably existing in the core of NSs.

In recent years, significant progresses have been made in observing and obtaining information about the properties of NS, such as mass, radius and tidal deformability.
A noteworthy event is the observation of the most accurate massive NS,
PSR J0740+6620~\cite{Cromartie2020,Fonseca2021},
which has a lower limit of $2.14_{-0.09}^{+0.10}~M_\odot$ (68.3\% credibility interval).
This measurement rules out those EOS which predict NS maximum mass ($M_\mathrm{max}$) lower than this value.
Another remarkable discovery is the heaviest NS known as PSR J0952-0607,
which reports a mass of $2.35\pm{0.17}~M_\odot$~\cite{Romani2022} (68.3\% credibility interval).
Additional observations of massive NS include the accurate mass determination of
object like PSR J1614-2230 ($1.908\pm0.016M_\odot$)~\cite{Arzoumanian2018}
and PSR J0348+0432 ($2.01\pm0.04M_\odot$)~\cite{Antoniadis2013}.
The simultaneous measurements on mass and radius of PSR J0030+0451 from the Neutron Star Interior Composition Explorer (NICER) give the range (68\% credible interval) $M=1.34_{-0.16}^{+0.15}~M_{\odot}$, $R=12.71_{-1.19}^{+1.14}~\mathrm{km}$~\cite{Riley2019}
as well as $M=1.44_{-0.14}^{+0.15}~M_{\odot}$, $R=13.02_{-1.06}^{+1.24}~\mathrm{km}$~\cite{Miller2019}.
Similar measurements from NICER for PSR J0740+6620 provide the results (68.3\% credibility interval) $M=2.072_{-0.066}^{+0.067}M_\odot$, $R=12.39_{-0.98}^{+1.30}~\mathrm{km}$~\cite{Riley2021} and $M=2.08\pm0.07M_\odot$, $R=13.7_{-1.5}^{+2.6}~\mathrm{km}$~\cite{Miller2021}.
The binary NS merger event GW170817 provided a constraint on the tidal deformability of a canonical $1.4~M_\odot$ NS with $\Lambda_{1.4}<800$~\cite{Abbott2017}.
Furthermore, the inferred constraint on $R_{1.4}$  from GW170817 indicates $R_{1.4}<13.6~\rm{km}$~\cite{Annala2018}.
In another gravitational wave event, the secondary component of GW190814 with a mass of $2.5\sim2.67~M_\odot$~\cite{Abbott2020} could potentially be a NS.
In this case, the corresponding constraint on tidal deformability reads $458\leq\Lambda_{1.4}\leq889$~\cite{Abbott2020}.

In the present work, we aim to investigate the impact of a new degree of freedom in NS matter, namely the H particle.
We explore the possible coupling strengths of H particle as a free parameter for the effective interaction within
the relativistic mean-field (RMF) model.
We treat the isovector coupling by two methods: (1) using a constant isovector coupling referred to as RMF model;
and 
(2) using a density-dependent isovector coupling referred to as RMFL model.
In the RMFL model, the density dependence introduced in the isovector $\rho$-meson channel follows the same functional form as in the density-dependent RMF (DDRMF) model~\cite{Spinella2017}.
In the general framework of covariant density functional (CDF) theory with density-dependent couplings, the density dependence is not limited to the isovector $\rho$ meson; the $\sigma$ and $\omega$ nucleon couplings also acquire explicit density dependence through phenomenological functions, as summarized in Ref.~\cite{Sedrakian2023}.
Furthermore, we analyze the effects of H particle on the properties of NS.
Before proceeding to the hadronic-level analysis, we compute the mass of the H particle within the chromomagnetic interaction model~\cite{Liu:2019zoy},  in which quarks interact via color forces.

The manuscript is organized in the following way.
The Chromomagnetic Interaction (CMI) framework is presented in Section~\ref{sec2} to calculate the mass of H particle.
In Section~\ref{sec:RMF} we describe NS matter containing H particle using RMF/RMFL models.
Our main results concerning the appearance and effects of H particle on NSs are shown and discussed in Section~\ref{sec:results}
Finally, the conclusions are provided at the end of the manuscript in Section~\ref{sec:conclusions}.
%

\section{mass spectrum of $nnnnss$ state}\label{sec2}
\subsection{Effective Hamiltonian and corresponding parameters}
For the ground mass of $nnnnss$ $(I=0)$ states,
we calculate it using the CMI model \cite{Weng:2021ngd,Liu:2019zoy,Weng:2019ynv,Liu:2023rzu,An:2021vwi,Weng:2020jao}.
Here, the $n$ quark in $nnnnss$ states represents the $u$ or $d$ quark.
As for the CMI model, it can well reproduce the masses of traditional mesons and baryons \cite{Liu:2019zoy}.
For instance, in the field of double-charm baryons,
Karliner et al. predicted the mass of $\Xi_{cc}$ to be $3627\pm12$ MeV within the CMI model framework~\cite{Karliner:2014gca}.
In 2017, the LHCb collaboration discovered the double-charm baryon $\Xi^{++}_{cc}$ ($ccu$) in the invariant mass spectrum of $\Lambda^{+}_{c}K^{-}\pi^{+}\pi^{-}$ from $pp$ collision experiments, with the mass of this state is $m_{\Xi^{++}_{cc}}=3621.4$ MeV \cite{LHCb:2017iph}.
The difference between this experimental value and the theoretical prediction in the framework of the CMI model is less than 10 MeV, confirming the model's reliability and predictive power.
Furthermore, inspired by the observation of $\Xi_{cc}^{++}(3621)$, Karliner and Eichten independently studied tetraquarks with the $QQ\bar{q}\bar{q}$ configuration (where $Q$ denotes a heavy quark)~\cite{Karliner:2017qjm, Eichten:2017ffp}.
Both studies are based on the mass spectrum of $QQ\bar{q}\bar{q}$ tetraquarks calculated from the CMI model \cite{Luo:2017eub},
and they further calculated and predicted an ``ideal" stable state $T_{bb\bar{u}\bar{d}}$ - a state that cannot decay through strong or electromagnetic interactions, but only via weak interactions to final states.
This is the first exotic hadron with such a property, and it also highlights the important role of the CMI model in hadron physics research \cite{Liu:2019zoy}.
Therefore, it is natural to extend this model to exotic hadrons (tetraquarks, pentaquarks, and hexaquarks).
The corresponding effective Hamiltonian for the CMI model is:
\begin{eqnarray}\label{Eq3}
H&=&\sum_im^{0}_i+H_{\rm CEI}+H_{\rm CMI}\nonumber\\
&=&\sum_im^{0}_i-\sum_{i<j}A_{ij} \vec\lambda_i\cdot \vec\lambda_j-\sum_{i<j}v_{ij} \vec\lambda_i\cdot \vec\lambda_j \vec\sigma_i\cdot\vec\sigma_j, \nonumber\\
&=&-\frac{3}{4}(-\frac{4}{3}\sum_im^{0}_i)-\sum_{i<j}A_{ij} \vec\lambda_i\cdot \vec\lambda_j\nonumber\\&&-\sum_{i<j}v_{ij} \vec\lambda_i\cdot \vec\lambda_j \vec\sigma_i\cdot\vec\sigma_j, \nonumber\\
&=&-\frac{3}{4}\sum_{i<j}[\frac{1}{4}(m^{0}_{i}+m^{0}_{j})+\frac{4}{3}A_{ij}] \vec\lambda_i\cdot \vec\lambda_j\nonumber\\&&-\sum_{i<j}v_{ij} \vec\lambda_i\cdot \vec\lambda_j \vec\sigma_i\cdot\vec\sigma_j, \nonumber\\
&=&-\frac{3}{4}\sum_{i<j}m_{ij}V^{\rm C}_{ij}-\sum_{i<j}v_{ij}V^{\rm CMI}_{ij},
\end{eqnarray}
where $V^{\rm C}_{ij}=\vec\lambda_i\cdot \vec\lambda_j$ and $V^{\rm CMI}_{ij}=\vec\lambda_i\cdot \vec\lambda_j \vec\sigma_i\cdot\vec\sigma_j$ represent the color interaction and chromomagnetic interaction between quarks, respectively.
Here, $\sigma_{i}$ denotes the Pauli matrix and $\lambda_i$ is the Gell-Mann matrix.
In the above expression, we have used the following formula derivation:
\begin{eqnarray}
&&-\frac{3}{4}\sum_{i<j}[\frac{1}{4}(m^{0}_{i}+m^{0}_{j})+\frac{4}{3}A_{ij}]\vec\lambda_i\cdot \vec\lambda_j\nonumber\\
&=&-\frac{3}{4}\times\frac{1}{4}\times\sum_{i<j}(m^{0}_{i}+m^{0}_{j})\vec\lambda_i\cdot \vec\lambda_j\nonumber-\sum_{i<j}A_{ij}\vec\lambda_i\cdot \vec\lambda_j\nonumber\\
&=&-\frac{3}{4}\times\frac{1}{4}\times[\frac{1}{2}\sum_{i,j}(m^{0}_{i}+m^{0}_{j})\vec\lambda_i\cdot \vec\lambda_j-\sum_{i}m^{0}_{i}(\vec\lambda_i)^{2}]\nonumber\\
&&-\sum_{i<j}A_{ij}\vec\lambda_i\cdot \vec\lambda_j\nonumber\\
&=&-\frac{3}{4}\times\frac{1}{4}\times[(\sum_{i}m^{0}_{i}\vec\lambda_i)\cdot(\sum_{i}\vec\lambda_i)-\frac{16}{3}\sum_{i}m^{0}_{i}]\nonumber\\
&&-\sum_{i<j}A_{ij}\vec\lambda_i\cdot \vec\lambda_j\nonumber\\
&=&\sum_im^{0}_i-\sum_{i<j}A_{ij} \vec\lambda_i\cdot \vec\lambda_j.
\end{eqnarray}
In this derivation, the color operator $(\sum_{i}\vec\lambda_i)$ equals zero in colorless physical states, so this term is omitted {in the final step}.
Furthermore, we can introduce a new mass parameter $m_{ij}$ for quark pairs, which is defined as:
\begin{eqnarray}
m_{ij}=\frac{1}{4}(m^{0}_{i}+m^{0}_{j})+\frac{4}{3}A_{ij},
\end{eqnarray}
where $m^{0}_{i}$ and $A_{ij}$ denote the effective mass of the $i$-th constituent quark (including the constituent quark mass and kinetic energy) and the strength of the chromoelectric interaction between the $i$-th and $j$-th quarks, respectively.
Parameter $m_{ij}$ mainly determines the size of the mass.
$v_{ij}$ is the effective coupling constant between the $i$-th quark and $j$-th quark and it mainly governs the mass gaps.
It depends on the quark masses and the spatial wave function of the ground-state baryon.
Using these conventional hadron masses, we {can} fix the coupling parameters $m_{ij}$ and $v_{ij}$ (see Table \ref{parameter2}).
Interested readers may further refer to Refs.~\cite{Weng:2018mmf,Weng:2021hje,An:2020vku,Liu:2021gva,Liu:2023rzu} for more details.

\begin{table*}[t]
\centering \caption{
Determination of coupling parameters for the $nnnnss$ state: inputted hadron masses and outputted coupling parameters (in units of MeV).
}\label{parameter2}
\renewcommand\arraystretch{2.45}
\renewcommand\tabcolsep{8pt}
\begin{tabular}{c|ccc|ccccc}
\bottomrule[1.5pt]
\bottomrule[0.5pt]
Hadron&$N$&\multicolumn{1}{c|}{$\Delta$}&\multicolumn{1}{c}{$\Lambda$}&$\Sigma$&\multicolumn{1}{c|}{$\Sigma^{*}$}&$\Xi$&\multicolumn{1}{c|}{$\Xi^{*}$}&$\Omega$\\
\bottomrule[0.5pt]
$I(J^{P})$&$1/2(1/2^{+})$&\multicolumn{1}{c|}{$3/2(3/2^{+})$}&\multicolumn{1}{c}{$0(1/2^{+})$}&\multicolumn{1}{c}{$1(1/2^{+})$}&\multicolumn{1}{c|}{$1(3/2^{+})$}&\multicolumn{1}{c}{$1/2(1/2^{+})$}&\multicolumn{1}{c|}{$1/2(3/2^{+})$}&\multicolumn{1}{c}{$0(3/2^{+})$}\\
\bottomrule[0.5pt]
Mass&938.3&\multicolumn{1}{c|}{1232.0}&\multicolumn{1}{c}{1115.7}&1189.4&\multicolumn{1}{c|}{1382.8}&1314.9&\multicolumn{1}{c|}{1531.8}&1672.5\\
\bottomrule[1pt]
Parameter&$m_{nn}$&$m_{ns}$&$m_{ss}$&$v_{nn}$&$v_{ns}$&\multicolumn{1}{c|}{$v_{ss}$}\\
\Xcline{1-7}{0.5pt}
Value&181.2&226.7&262.3&19.1&13.3&\multicolumn{1}{c|}{12.2}\\
\bottomrule[0.5pt]
\bottomrule[1.5pt]
\end{tabular}
\end{table*}


\subsection{Wave Functions}\label{sec3}

To calculate the mass spectrum of $nnnnss$ ($I=0$) state, we also need to construct the total wave function that satisfied the Pauli principle.
The total wave function of $nnnnss$ ($I=0$) state is described by the direct product of spatial, flavor, color, and spin wave functions
\begin{eqnarray}
\psi_{\textrm{tot}}=\psi_{space}\otimes\psi_{flavor}\otimes\psi_{color}\otimes\psi_{spin}.
\end{eqnarray}
Here, we only consider low-lying
$S$-wave hexaquark states, {since} the symmetry constraint from the spatial wave function is trivial.
Thus, the remaining components-ie. the $\psi_{flavor}\otimes\psi_{color}\otimes\psi_{spin}$ components- must be fully antisymmetric under the exchange of identical quarks.
For $nnnnss$ ($I=0$) state, we need to construct $\psi_{flavor}\otimes\psi_{color}\otimes\psi_{spin}$ wave functions {satisfying} $\{1234\}\{56\}$ symmetry.
Here, we use the notation $\{1234\}$ ($\{56\}$) to indicate that quarks 1, 2, 3, and 4 (5 and 6) are antisymmetric under exchange;
similarly, we also use the notation $[1234]$ ($[56]$) to indicate that quarks 1, 2, 3, and 4 (quarks 5 and 6) are symmetric under exchange.

Moreover, Young tableaux, which {represent} the irreducible bases of the permutation group,
enable us to easily identify the hexaquark configuration with certain symmetry property \cite{An:2021vwi,Park:2017jbn,Park:2015nha,An:2020jix,Park:2016cmg,Park:2016mez}.
Therefore, we could use the Young diagram, Young tableaux, and Young-Yamanouchi basis vectors to describe these wave functions.
\subsubsection{Flavor part}\label{1}

For the flavor part of the $nnnnss$ ($I=0$) state,
we consider $SU(2)$ flavor symmetry and {treat} the $s$ quark as a flavor singlet, justified by the mass of $s$ quark is much heavier than those of the $u$ and $d$ quarks.
When $SU(2)$ symmetry holds,
and {the flavor wave functions of the $s$ quark can be decoupled from those of the $u$, $d$ quarks,}
the $s$ quark can be effectively treated as a flavor singlet.
The corresponding Young diagram of the $nnnnss$ ($I=0$) state is shown in:
\begin{align}\label{f}
(\begin{tabular}{|c|c|c|}
\hline
   $\quad$&$\quad$ \\
   \cline{1-2}
     $\quad$&$\quad$  \\
   \cline{1-2}
\end{tabular}_{I=0}
;
\begin{tabular}{|c|c|c|}
\hline
    $\quad$  & $\quad$ \\
   \cline{1-2}
\end{tabular})_{F}.
\end{align}

Based on this Young diagram, we derive two corresponding Young tableaux:

\begin{align}
(\begin{tabular}{|c|c|c|}
\hline
   1&2 \\
\cline{1-2}
   3&4  \\
\cline{1-2}
\end{tabular}_{I=0}
;
\begin{tabular}{|c|c|c|}
\hline
5 & 6 \\
\cline{1-2}
\end{tabular})_{F};\quad
(\begin{tabular}{|c|c|c|}
\hline
   1&3 \\
\cline{1-2}
   2&4  \\
\cline{1-2}
\end{tabular}_{I=0}
;
\begin{tabular}{|c|c|c|}
\hline
5 & 6 \\
\cline{1-2}
\end{tabular})_{F}.
\end{align}


\subsubsection{Color part}

For the color part, we only need to consider color singlets based on color confinement, and this part satisfies
$SU(3)$ symmetry.
We first construct the Young diagram corresponding to the color singlet of the hexaquark state:
\begin{align}\label{color6}
\begin{tabular}{|c|c|}
\hline
 $\quad$ &  $\quad$   \\
\cline{1-2}
 $\quad$ &  $\quad$   \\
\cline{1-2}
 $\quad$ &  $\quad$   \\
\cline{1-2}
\end{tabular}_{C}.
\end{align}

Based on this Young diagram (Eq.(\ref{color6})), we {derive} five corresponding Young tableaux {that represent} color singlets:
\begin{align}\label{eq-color1}
C_{1}=\begin{tabular}{|c|c|}
\hline
1 &  2   \\
\cline{1-2}
3 & 5 \\
\cline{1-2}
4 & 6  \\
\cline{1-2}
\end{tabular},
C_{2}=\begin{tabular}{|c|c|}
\hline
1 &  2   \\
\cline{1-2}
3 & 4 \\
\cline{1-2}
5 & 6  \\
\cline{1-2}
\end{tabular},
C_{3}=\begin{tabular}{|c|c|}
\hline
1 & 3 \\
\cline{1-2}
2 & 4 \\
\cline{1-2}
5 & 6  \\
\cline{1-2}
\end{tabular},
C_{4}=\begin{tabular}{|c|c|}
\hline
1 & 3 \\
\cline{1-2}
2 & 5 \\
\cline{1-2}
4 & 6  \\
\cline{1-2}
\end{tabular},
C_{5}=\begin{tabular}{|c|c|}
\hline
1 & 4 \\
\cline{1-2}
2 & 5 \\
\cline{1-2}
3 & 6  \\
\cline{1-2}
\end{tabular}.
\end{align}

Based on the above Young tableaux (Eq.(\ref{eq-color1})), {we split them according to the [1234] and [56] grouping.
This allows us to obtain the following Young diagram:}
\begin{align}\label{color}
(\begin{tabular}{|c|c|c|}
\hline
 $\quad$ &  $\quad$   \\
\cline{1-2}
 $\quad$ &  $\quad$   \\
\cline{1-2}
\end{tabular}
;
\begin{tabular}{|c|c|c|}
\hline
 $\quad$ &  $\quad$   \\
\hline
\end{tabular})_{C}, \quad
(\begin{tabular}{|c|c|c|}
\hline
 $\quad$ &  $\quad$   \\
\cline{1-2}
 $\quad$   \\
\cline{1-1}
 $\quad$   \\
\cline{1-1}
\end{tabular}
;
\begin{tabular}{|c|c|c|}
\hline
 $\quad$  \\
\hline
 $\quad$  \\
\hline
\end{tabular})_{C}.
\end{align}

\subsubsection{Spin part}
The spin part {is} based on the $SU(2)$ symmetry.
For $nnnnss$ states, all possible total {spins} are $J=$ $3, 2, 1$, and $0$, respectively.
For these spin states, we can {construct} four corresponding Young diagram:
\begin{align}\label{spin}
\begin{tabular}{|c|c|c|c|c|c|}
\hline
 $\quad$ & $\quad$ & $\quad$& $\quad$ & $\quad$& $\quad$\\
\cline{1-6}
\end{tabular}_{S},
\quad
&\begin{tabular}{|c|c|c|c|c|}
\hline
 $\quad$ & $\quad$ & $\quad$& $\quad$ & $\quad$\\
\cline{1-5}
 $\quad$  \\
\cline{1-1}
\end{tabular}_{S},
\nonumber\\
\begin{tabular}{|c|c|c|c|}
\hline
 $\quad$ & $\quad$ & $\quad$& $\quad$ \\
\cline{1-4}
 $\quad$ & $\quad$  \\
\cline{1-2}
\end{tabular}_{S},\quad
&\begin{tabular}{|c|c|c|}
\hline
 $\quad$ & $\quad$ & $\quad$ \\
\cline{1-3}
 $\quad$ & $\quad$ & $\quad$ \\
\cline{1-3}
\end{tabular}_{S}.
\end{align}

{Using} these Young diagrams (Eq.(\ref{spin})), we {derive} twenty corresponding Young tableaux {that correspond to} different spin states:

\begin{equation}\label{eq-color1}
\begin{split}
J=3:&\begin{tabular}{|c|c|c|c|c|c|}
\hline
1 &  2 &3&4&5 &6  \\
\cline{1-6}
\end{tabular}_{1};\\
J=2:
&\begin{tabular}{|c|c|c|c|c|c|}
\hline
1 &  2 &3&4&5   \\
\cline{1-5}
6\\
\cline{1-1}
\end{tabular}_{2},
\begin{tabular}{|c|c|c|c|c|}
\hline
1 &  2 &3&4&6   \\
\cline{1-5}
5\\
\cline{1-1}
\end{tabular}_{3},
\begin{tabular}{|c|c|c|c|c|}
\hline
1 &  2 &3&5 &6  \\
\cline{1-5}
4\\
\cline{1-1}
\end{tabular}_{4},
\\
&\begin{tabular}{|c|c|c|c|c|}
\hline
1 &  2 &4&5 &6  \\
\cline{1-5}
3\\
\cline{1-1}
\end{tabular}_{5},
\begin{tabular}{|c|c|c|c|c|}
\hline
1 &  3 &4&5 &6  \\
\cline{1-5}
2\\
\cline{1-1}
\end{tabular}_{6};\\
J=1:&
\begin{tabular}{|c|c|c|c|c|}
\hline
1 &  2 &3&4   \\
\cline{1-4}
5&6\\
\cline{1-2}
\end{tabular}_{7},\quad
\begin{tabular}{|c|c|c|c|c|}
\hline
1 &  2 &3&5   \\
\cline{1-4}
4&6\\
\cline{1-2}
\end{tabular}_{8},
\begin{tabular}{|c|c|c|c|c|}
\hline
1 &  2 &3 &6  \\
\cline{1-4}
4&5\\
\cline{1-2}
\end{tabular}_{9},
\\
&
\begin{tabular}{|c|c|c|c|c|}
\hline
1 &  2 &4&5   \\
\cline{1-4}
3&6\\
\cline{1-2}
\end{tabular}_{10},
\begin{tabular}{|c|c|c|c|c|}
\hline
1 &  2 &4&6   \\
\cline{1-4}
3&5\\
\cline{1-2}
\end{tabular}_{11},
\begin{tabular}{|c|c|c|c|c|}
\hline
1 &  2 &5&6   \\
\cline{1-4}
3&4\\
\cline{1-2}
\end{tabular}_{12},
\\
&
\begin{tabular}{|c|c|c|c|c|}
\hline
1 &  3 &4&5   \\
\cline{1-4}
2&6\\
\cline{1-2}
\end{tabular}_{13},
\begin{tabular}{|c|c|c|c|c|}
\hline
1 &  3 &4&6  \\
\cline{1-4}
2&5\\
\cline{1-2}
\end{tabular}_{14},
\begin{tabular}{|c|c|c|c|c|}
\hline
1 & 3 &5&6  \\
\cline{1-4}
2&4\\
\cline{1-2}
\end{tabular}_{15};\\
J=0:&
\begin{tabular}{|c|c|c|c|c|}
\hline
1 &  2 &3   \\
\cline{1-3}
4&5&6\\
\cline{1-3}
\end{tabular}_{16},
\begin{tabular}{|c|c|c|c|c|}
\hline
1 &  2 &4   \\
\cline{1-3}
3&5&6\\
\cline{1-3}
\end{tabular}_{17},
\begin{tabular}{|c|c|c|c|c|}
\hline
1 & 2 &5  \\
\cline{1-3}
3&4&6\\
\cline{1-3}
\end{tabular}_{18},
\begin{tabular}{|c|c|c|c|c|}
\hline
1 &  3 &4   \\
\cline{1-3}
2&5&6\\
\cline{1-3}
\end{tabular}_{19},
\begin{tabular}{|c|c|c|c|c|}
\hline
1 &  3 &5  \\
\cline{1-3}
2&4&6\\
\cline{1-3}
\end{tabular}_{20}.
\end{split}
\end{equation}

\subsubsection{Flavor $\otimes$ Color $\otimes$ Spin part}
With the above {preparations}, we can begin to construct
$\psi_{flavor}\otimes\psi_{color}\otimes\psi_{spin}$ wave functions {satisfying} $\{1234\}\{56\}$ symmetry.
For flavor $\otimes$ color $\otimes$ spin {sector, this part satisfies} $SU(12)$ symmetry.
Here, we {employ the reduction of $SU(12)_{FCS}\supset SU(2)_{F} \otimes SU(6)_{CS}$, expressed as}:

\begin{align}\label{fcs1}
(\begin{tabular}{|c|c|c|}
\hline
$\quad$ \\
\cline{1-1}
$\quad$ \\
\cline{1-1}
$\quad$ \\
\cline{1-1}
$\quad$ \\
\cline{1-1}
\end{tabular}
;
\begin{tabular}{|c|c|c|}
\hline
$\quad$ \\
\cline{1-1}
$\quad$ \\
\cline{1-1}
\end{tabular})_{FCS}
=
(\begin{tabular}{|c|c|c|}
\hline
   $\quad$&$\quad$ \\
   \cline{1-2}
     $\quad$&$\quad$  \\
   \cline{1-2}
\end{tabular}
;
\begin{tabular}{|c|c|c|}
\hline
    $\quad$  & $\quad$ \\
   \cline{1-2}
\end{tabular})_{F}
\otimes
(\begin{tabular}{|c|c|c|}
\hline
   $\quad$&$\quad$ \\
   \cline{1-2}
     $\quad$&$\quad$  \\
   \cline{1-2}
\end{tabular}
;
\begin{tabular}{|c|c|c|}
\hline
 $\quad$  \\
 \cline{1-1}
  $\quad$  \\
 \cline{1-1}
\end{tabular})_{CS}
\oplus...\nonumber\\
\end{align}

Based on Eq.(\ref{fcs1}), we can {derive} corresponding Young diagram of color $\otimes$ spin {sector} from the Young diagram of flavor {sector} (Eq.(\ref{f})):

\begin{align}
(\begin{tabular}{|c|c|c|}
\hline
   $\quad$&$\quad$ \\
   \cline{1-2}
     $\quad$&$\quad$  \\
   \cline{1-2}
\end{tabular}_{I=0}
;
\begin{tabular}{|c|c|c|}
\hline
    $\quad$  & $\quad$ \\
   \cline{1-2}
\end{tabular})_{F}
\Rightarrow
(\begin{tabular}{|c|c|c|}
\hline
   $\quad$&$\quad$ \\
   \cline{1-2}
     $\quad$&$\quad$  \\
   \cline{1-2}
\end{tabular}
;
\begin{tabular}{|c|c|c|}
\hline
 $\quad$  \\
 \cline{1-1}
  $\quad$  \\
 \cline{1-1}
\end{tabular})_{CS}.
\end{align}
\\

For color $\otimes$ spin {sector, this part satisfies} $SU(6)$ symmetry.
Here, {we can employ the reduction $SU(6)_{CS}$ to $SU(3)_{C}\otimes SU(2)_{S}$, expressed as:}
\begin{align}\label{cs}
&\quad(\begin{tabular}{|c|c|c|}
\hline
   $\quad$&$\quad$ \\
   \cline{1-2}
     $\quad$&$\quad$  \\
   \cline{1-2}
\end{tabular}
;
\begin{tabular}{|c|c|c|}
\hline
 $\quad$  \\
 \cline{1-1}
  $\quad$  \\
 \cline{1-1}
\end{tabular})_{CS}
\nonumber
\\
&=[(\begin{tabular}{|c|c|c|}
\hline
   $\quad$&$\quad$ \\
   \cline{1-2}
     $\quad$&$\quad$  \\
   \cline{1-2}
\end{tabular}_{C};
\begin{tabular}{|c|c|c|c|}
\hline
$\quad$&$\quad$&$\quad$&$\quad$ \\
\cline{1-4}
\end{tabular}_{S})
(\begin{tabular}{|c|c|c|}
\hline
   $\quad$&$\quad$ \\
   \cline{1-2}
\end{tabular}_{C};
\begin{tabular}{|c|c|c|c|}
\hline
$\quad$ \\
\cline{1-1}
$\quad$ \\
\cline{1-1}
\end{tabular}_{S})]
\nonumber \\
&\oplus[(\begin{tabular}{|c|c|c|}
\hline
   $\quad$&$\quad$ \\
   \cline{1-2}
     $\quad$&$\quad$  \\
   \cline{1-2}
\end{tabular}_{C};
\begin{tabular}{|c|c|c|}
\hline
   $\quad$&$\quad$ \\
   \cline{1-2}
     $\quad$&$\quad$  \\
   \cline{1-2}
\end{tabular}_{S})
(\begin{tabular}{|c|c|c|}
\hline
   $\quad$&$\quad$ \\
   \cline{1-2}
\end{tabular}_{C};
\begin{tabular}{|c|c|c|c|}
\hline
$\quad$ \\
\cline{1-1}
$\quad$ \\
\cline{1-1}
\end{tabular}_{S})] \nonumber \\
&\oplus
[(\begin{tabular}{|c|c|c|}
\hline
   $\quad$&$\quad$ \\
   \cline{1-2}
     $\quad$  \\
   \cline{1-1}
    $\quad$  \\
   \cline{1-1}
\end{tabular}_{C};
\begin{tabular}{|c|c|c|}
\hline
   $\quad$&$\quad$ &$\quad$\\
   \cline{1-3}
     $\quad$  \\
   \cline{1-1}
\end{tabular}_{S})
(\begin{tabular}{|c|c|c|c|}
\hline
$\quad$ \\
\cline{1-1}
$\quad$ \\
\cline{1-1}
\end{tabular}_{C};
\begin{tabular}{|c|c|c|}
\hline
   $\quad$&$\quad$ \\
   \cline{1-2}
\end{tabular}_{S})]...
\end{align}
{We can easily observe that these three Young diagrams for the color sector correspond exactly to the Young diagrams in Eq. (\ref{color}).
Further analysis shows that the first Young diagram for the spin sector corresponds to a $J=2$ state.
Similarly, the second Young diagram for the spin sector corresponds to a $J=0$ state,} whereas the third gives rise to $J=0, 1$ and 2 states.
Other Young diagrams about color part can not form the color {singlets and thus} need not be considered.
Based on the above discussions, {the ground $nnnnss$ $(I=0)$ state has two $J=2$ states, one $J=1$ state, and two $J=0$ states.}

{Using} the Clebsch-Gordan (CG) {coefficients} of the permutation group $S_{n}$, we obtain the coupling scheme designed to construct the color $\otimes$ spin wave functions.
The expression {for the} CG {coefficients of $S_{n}$  permutation group is given by} \cite{Stancu:1999qr,Park:2016cmg}
\begin{align}
&\quad S([f^{\prime}]p^{\prime}q^{\prime}y^{\prime}[f^{\prime\prime}]p^{\prime\prime}q^{\prime\prime}y^{\prime\prime}\vert[f]pqy)\nonumber\\
&=
K([f^{\prime}]p^{\prime}[f^{\prime\prime}]p^{\prime\prime}\vert[f]p)
S([f^{\prime}_{p^{\prime}}]q^{\prime}y^{\prime}[f^{\prime\prime}_{p^{\prime\prime}}]q^{\prime\prime}y^{\prime\prime}\vert[f_p]qy).
\label{K-matrix}
\end{align}
Here, $S$ in the left-hand (right-hand) side denotes a CG coefficient of $S_n$ ($S_{n-1}$) permutation group, and $K$ is an isoscalar factor, which is called $K$ matrix that factorizes the CG coefficients of $S_n$ into a CG coefficients of $S_{n-1}$ multiplied by the isoscalar factor.
In this notation, $[f]$ ($[f_p]$) is a Young tableau of the $S_{n}$ ($S_{n-1}$) permutation group, with $[f]pqy$ representing a specific Young-Yamanouchi basis vector.
$p$ ($q$) indicates the row of the $n$ $(n-1)$-th particle in the Young-Yamanouchi basis vector, and $y$ describes the distribution of the remaining $n-2$ particles.
{Using} the isoscalar factors for $S_{3}$ and $S_{4}$ in Tables 6.2 and 6.3 of Ref.\cite{Stancu:1999qr}, we can derive the corresponding CG coefficients for $S_{4}$ in Eq. (\ref{K-matrix}).

Based on the above, we obtain the Young-Yamanouchi basis {vectors} for the flavor $\otimes$ color $\otimes$ spin {sector}, which are derived from the relevant Young diagram equation (Eq.(\ref{fcs1})):


\begin{align}\label{fcs}
(\begin{tabular}{|c|c|c|}
\hline
1  \\
\cline{1-1}
2  \\
\cline{1-1}
3  \\
\cline{1-1}
4  \\
\cline{1-1}
\end{tabular}
;
\begin{tabular}{|c|c|c|}
\hline
5  \\
 \cline{1-1}
6  \\
 \cline{1-1}
\end{tabular})_{FCS}&=
\sqrt{\frac{1}{2}}(\begin{tabular}{|c|c|c|}
\hline
   1&2 \\
   \cline{1-2}
   3&4  \\
   \cline{1-2}
\end{tabular}_{I}
;
\begin{tabular}{|c|c|c|}
\hline
5&6  \\
 \cline{1-2}
\end{tabular})_{F}
\otimes
(\begin{tabular}{|c|c|c|}
\hline
1&3 \\
\cline{1-2}
2&4  \\
\cline{1-2}
\end{tabular}
;
\begin{tabular}{|c|c|c|}
\hline
5  \\
 \cline{1-1}
6 \\
 \cline{1-1}
\end{tabular})_{CS}
\nonumber\\
&-\sqrt{\frac{1}{2}}(\begin{tabular}{|c|c|c|}
\hline
   1&3 \\
   \cline{1-2}
   2&4  \\
   \cline{1-2}
\end{tabular}_{I}
;
\begin{tabular}{|c|c|c|}
\hline
5&6  \\
 \cline{1-2}
\end{tabular})_{F}
\otimes
(\begin{tabular}{|c|c|c|}
\hline
1&2 \\
\cline{1-2}
3&4  \\
\cline{1-2}
\end{tabular}
;
\begin{tabular}{|c|c|c|}
\hline
5  \\
 \cline{1-1}
6 \\
 \cline{1-1}
\end{tabular})_{CS}.
\end{align}

{Similarly, we derive the Young-Yamanouchi basis vectors for the color $\otimes$ spin sector from the relevant Young diagram equation (Eq.(\ref{cs})), which are presented in the Appendix \ref{sec10}.}

By combining the Young-Yamanouchi basis vectors in Eqs.(\ref{eq:jp015}-\ref{eq:jp025}) and Eq.(\ref{fcs}),
we {construct} all possible $\psi_{flavor}\otimes\psi_{color}\otimes\psi_{spin}$ wave functions {satisfying the} $\{1234\}\{56\}$ symmetry.
{Using} these wave functions,
we can calculate the CMI model Hamiltonian, and further {derive} the mass spectra of the $nnnnss$ $(I=0)$ state.

\subsection{Mass spectra results}\label{sec4}
The mass spectra of $nnnnss$ ($I=0$) state {are} shown in Table~\ref{tab:2Hmass}, in which the $J^{P}=0^{+}$ state has $m_H=m_{uuddss}=2212.7~\rm{MeV}$.

{Since} the mass $m_H$ of {the H particle} is lower than the $\Lambda \Lambda$ threshold, {the} H particle could be an ideal stable {constituent} inside neutron stars.
According to {Eq.~(\ref{eq:jp015}) and Eq.~(\ref{eq:jp021})}, there are two wave functions satisfying Pauli principle for the {$J^{P}=0^{+}$ states}.
{Thus, the Hamiltonian corresponding to} Eq.~(\ref{Eq3}) {forms} a $2\times2$ matrix;
the eigenvalues of Eq.~(\ref{Eq3}) indicates that there are two states in the $0^{+}$ state, between which there is flavor mixing.
Meanwhile, their mass splitting between them is determined by both the chromomagnetic interaction and chromoelectric interaction. 
\begin{table}[h]
\centering \caption{The mass of the $nnnnss$ ($I=0$) state (in units of \rm{MeV}).
}\label{mass}\label{tab:2Hmass}
\renewcommand\arraystretch{1.45}
\renewcommand\tabcolsep{7pt}
\begin{tabular}{c|cc|c|cc}
\bottomrule[1.5pt]
\bottomrule[0.5pt]
$J^{P}$&\multicolumn{2}{c|}{$0^{+}$}&$1^{+}$&\multicolumn{2}{c}{$2^{+}$}\\
\bottomrule[0.5pt]
Mass&2569.4&2212.7&2360.1&2752.3&2458.8\\
\bottomrule[0.5pt]
\bottomrule[1.5pt]
\end{tabular}
\end{table}

\section {The RMF/RMFL model}
\label{sec:RMF}
The RMF/RMFL model is constructed based on the meson-exchange picture which describes the nucleon-nucleon ($NN$) interaction.
In this model, the scalar ($\sigma$) meson represents the attractive component, while the vector meson ($\omega$) represents the repulsive component of the NN interaction.
Additionally, the isovector meson ($\rho$) is introduced to account for the isospin difference between neutron ($n$) and proton ($p$).
In this work, we focus on the properties of the H particle, which consists of the $uuddss$ quarks, including strangeness degree.
To achieved this, we calculate the rest mass of H particle ($m_H=2212.7~\rm{MeV}$) in Sec.~\ref{sec2}, the interaction between H particles in NS matter is transmitted through mesons, similar to nucleon interactions.
For simplicity, we do not consider the contribution of hidden-strangeness mesons ($\sigma^*$, $\phi$) here, which describe attractive and repulsive interactions involving strangeness.
Instead, we employ the free couplings of $g_{\sigma H}$ and $g_{\omega H}$ to characterize the attractive and repulsive interactions of the H particle,
in which, we effectively account for the attractive and repulsive contribution from strangeness mesons.
It is worth noting that information regarding the H particle matter system is limited,
and there is currently no consensus on the magnitude of the H particle potential.
Therefore, adopting this approach allows us to reduce the number of free parameters in our study.

We utilize the RMF/RMFL model as our theoretical framework.
The total Lagrangian consisting of nucleons, leptons, mesons and H particles is given by
\begin{eqnarray}
\mathcal{L}_{\rm{RMFL}} =\mathcal{L}_{\rm{N}}+\mathcal{L}_{\rm{l}}+\mathcal{L}_{\rm{m}}+\mathcal{L}_{\rm{H}} ,
\end{eqnarray}
where
\begin{eqnarray}
\mathcal{L}_{\rm{N}}& = & \sum_{i=p,n}\bar{\psi}_i
\bigg \{i\gamma_{\mu}\partial^{\mu}-M^{*}
\notag\\
&&-\gamma_{\mu} \left[g_{\omega}\omega^{\mu} +\frac{g_{\rho}}{2}\tau_a\rho^{a\mu}
\right]\bigg \}\psi_i  ,\\
\mathcal{L}_{\rm{l}}& = &\sum_{l=e,\mu}\bar{\psi}_{l}
  \left( i\gamma_{\mu }\partial^{\mu }-m_{l}\right)\psi_l, \\
\mathcal{L}_{\rm{m}}
& = &\frac{1}{2}\partial_{\mu}\sigma\partial^{\mu}\sigma -\frac{1}{2}%
m^2_{\sigma}\sigma^2-\frac{1}{3}g_{2}\sigma^{3} -\frac{1}{4}g_{3}\sigma^{4}
\notag \\
&& -\frac{1}{4}W_{\mu\nu}W^{\mu\nu} +\frac{1}{2}m^2_{\omega}\omega_{\mu}%
\omega^{\mu} +\frac{1}{4}c_{3}\left(\omega_{\mu}\omega^{\mu}\right)^2
\notag \\
&& -\frac{1}{4}R^a_{\mu\nu}R^{a\mu\nu} +\frac{1}{2}m^2_{\rho}\rho^a_{\mu}%
\rho^{a\mu}
\notag \\
&& +\Lambda_{\rm{v}}g^2_{\omega}g^2_{\rho} \omega_{\mu}%
\omega^{\mu} \rho^a_{\mu}\rho^{a\mu},
\\
\mathcal{L}_{\rm{H}}&=&\left(\partial_\mu-i g_{\omega H}\omega_\mu\right)\phi^{*}_H \left(\partial^\mu+i g_{\omega H}\omega^\mu\right)\phi_H
\notag \\
&&-\left(m_H+g_{\sigma H}\sigma\right)^2\phi^{*}_H\phi_H,
\end{eqnarray}
with $M^*=M+g_\sigma \sigma$ representing the effective mass of the nucleon.
In the context of homogeneous nuclear matter, the meson field equation takes the following forms,
\begin{eqnarray}
&&m_{\sigma }^{2}\sigma +g_{2}\sigma ^{2}+g_{3}\sigma
^{3}=-g_{\sigma }\left( n_{p}^{s}+n_{n}^{s}\right) - g_{\sigma H}n_{H} ,\\
&&m_{\omega }^{2}\omega +c_{3}\omega^{3}
+2\Lambda_{\rm{v}}g^2_{\omega}g^2_{\rho}{\rho}^2 \omega
=g_{\omega}\left( n_{p}+n_{n}\right) +g_{\omega H}n_{H} ,\notag\\
\\
&&m_{\rho }^{2}{\rho}
+2\Lambda_{\rm{v}}g^2_{\omega}g^2_{\rho}{\omega}^2{\rho}
=\frac{g_{\rho }}{2}\left(n_{p}-n_{n}\right)  ,
\end{eqnarray}
where $n_i^s$ and $n_i$ ($i=n, p$) represent the scalar density and vector density, respectively.
The isovector coupling $g_\rho$ in RMFL is treated as a density-dependent function, following the approach used in the density-dependent RMF (DDRMF) framework,
\begin{eqnarray}
g_{\rho}(n_b)=g_{\rho}(n_0)\exp\left[-a_{\rho}\left(\frac{n_b}{n_0}-1\right)\right],
\label{eq:g_r}
\end{eqnarray}%
where $n_0$ is the saturation density.
The density dependence of $g_{\rho}$ contributes a rearrangement term for the nucleon vector potential,
\begin{eqnarray}
\Sigma_{r}=\frac{1}{2}\sum_{i=p,n}\frac{\partial{g_{\rho}(n_b)}}{\partial{n_b}}\tau_3{n_i}{\rho}
=-\frac{1}{2}a_{\rho}g_{\rho}(n_b)\frac{n_p-n_n}{n_0}{\rho}. \nonumber\\
\label{eq:er}
\end{eqnarray}%
The chemical potential of proton, neutron, leptons, and H particle are given by
\begin{eqnarray}
\mu_{p} &=& {\sqrt{\left( k_{F}^{p}\right)^{2}+{M^{\ast }}^{2}}}+g_{\omega}\omega+\Sigma_{r}   +\frac{g_{\rho}(n_b)}{2}\rho,
\label{eq:mup} \\
\mu_{n} &=& {\sqrt{\left( k_{F}^{n}\right)^{2}+{M^{\ast }}^{2}}}+g_{\omega}\omega+\Sigma_{r}  -\frac{g_{\rho}(n_b)}{2}\rho,
\label{eq:mun} \\
\mu_{H}&=&m_{H}+g_{\sigma H}{\sigma}+g_{\omega H}{\omega},\\
\mu_l&=&\sqrt{k_{F_l}^2+m_l^2} ,~ (l=e,\mu).
\end{eqnarray}
The $\beta$ equilibrium conditions in NS require
\begin{eqnarray}
\mu_n &=& \mu_{p}+\mu_{e},\\
\mu_{\mu}&=&\mu_{e}, \\
\mu_H &=& 2\mu_{n}.
\label{eq:Hequlibrium}
\end{eqnarray}
The decay path $nn\rightarrow H \rightarrow nn$ is considered in this RMF framework, as shown in Eq.~(\ref{eq:Hequlibrium}).
In the H particle admixed NS matter, the total pressure and energy density are given by
\begin{eqnarray}
P &=& \sum_{i=p,n}\frac{1}{3\pi^2}\int_{0}^{k^{i}_{F}}
      \frac{1}{\sqrt{k^2+{M^{\ast}}^2}}k^4dk    \nonumber  \\
&& - \frac{1}{2}m^2_{\sigma}{\sigma}^2-\frac{1}{3}{g_2}{\sigma}^3
     -\frac{1}{4}{g_3}{\sigma}^4      + \frac{1}{2}m^2_{\omega}{\omega}^2
         \nonumber \\
      && +\frac{1}{4}{c_3}{\omega}^4+ \frac{1}{2}m^2_{\rho}{\rho}^2
      +n_b{\Sigma_{r}}+\Lambda_{\rm{v}}g^2_{\omega}g^2_{\rho} {\omega}^2{\rho}^2 \nonumber\\
      &&+\sum_{l=e,\mu}\frac{1}{3\pi^2}\int_0^{k_F^l}\frac{k^4 dk}{\sqrt{k^2+m^2_l}},
 \label{eq:php}\\
\varepsilon&=&\sum_{b=n,p} \frac{1}{\pi^2}\int_0^{k_{F_b}}\sqrt{k^2+m^{*2}_b}k^2 dk \notag\\
&&+ \frac{1}{2}m^2_{\sigma}{\sigma}^2+\frac{1}{3}{g_2}{\sigma}^3
     +\frac{1}{4}{g_3}{\sigma}^4
 + \frac{1}{2}m^2_{\omega}{\omega}^2 \nonumber  \\
 &&+\frac{3}{4}{c_3}{\omega}^4+\frac{1}{2}m_\rho^2{\rho}^2
+3\Lambda_{\rm{v}}g^2_{\omega}g^2_{\rho} {\omega}^2{\rho}^2+m^*_{H}n_{H}
\nonumber\\
&&+\sum_{l=e,\mu}\frac{1}{\pi^2}\int_0^{k_F^l}\sqrt{k^2+m^2_l}k^2 dk .
\label{eq:ehp}
\end{eqnarray}
It is important to highlight that, unlike fermions which could contribute to pressure via Fermi degeneracy,
the H particle, being a boson, influences the pressure only through its effect on the meson coupling potential.
When $a_\rho=0$, the rearrangement term $\Sigma_r$ vanishes.
As a result, the isovector coupling $g_\rho$ becomes a constant, and the RMFL model reduces to a general RMF model.

We note that H particle does not directly contribute to the pressure in NS matter.
However, by affecting the strength of meson field, the presence of H particle can still influence the total pressure of the system.
We apply four different parameter sets for the meson couplings, BigApple~\cite{Fattoyev2020},
NL3L-50~\cite{Wu2021,Ju2025}, TM1e~\cite{Bao2014,Wu2018} and NL3~\cite{Lalazissis1997}, which are listed in Table~\ref{tab:para}.
The existence of massive NSs requires a stiff EOS at the core.
However, constraints of the tidal deformability and supernova explosion mechanism favor a soft normal neutron star crust.
Among the parameter sets used in this study, BigApple, NL3L-50 and TM1e satisfy the maximum mass and tidal deformability requirements based on observation constraints.
The NL3 parameter set is included as a comparison.
The couplings between H particle and mesons are taken as free parameters in this study.
In contrast to the choice made in Ref.~\cite{Mantziris2020}, we consider only positive values for the dimensionless couplings $\chi_{\sigma H}=g_{\sigma H}/g_\sigma$ and $\chi_{\omega H}=g_{\omega H}/g_\omega$.
For example, with $\chi_{\sigma H}=0$, a purely repulsive interaction similar to the negative values in Ref.~\cite{Mantziris2020} is formed.
The same applies to $\chi_{\omega H}=0$ for the case of a purely attractive interaction.
As we lack information of H particle potential, we explore a wide range
for $\chi_{\sigma H}$ and $\chi_{\omega H}$ to determine the appropriate scale.
In Ref.~\cite{Mantziris2020}, the increasing fraction of $d^*(2380)$ leads to an increase in the fraction of electrons, which contributes to the pressure.
However, unlike $d^*(2380)$, the charge-neutral H particle does not have such an effect on electron fraction.
Therefore, the presence of the H particle has a more pronounced impact on reducing the pressure compared to $d^*(2380)$.
Usually a positive contribution of H particle to the pressure is favored when introducing a new degree of freedom at high densities in NS matter.
In our study, we focus on the constraints derived from observations
and entertain the possibility of H particle existing in a stable NS.

\begin{table}[htbp]
\caption{Masses of nucleons and mesons and meson coupling constants.
The masses are given in $\rm{MeV}$.}
\centering
\label{tab:para}
\setlength{\tabcolsep}{2.4mm}{
\begin{tabular}{lccccccc}
\hline\hline
Parameters   &BigApple  &NL3L-50  &TM1e  &NL3   \\
\hline
$M$            &939.0    &939.0    &938.0    &939.0 \\
$m_\sigma$     &492.730  &508.194  &511.198  &508.194 \\
$m_\omega$     &782.500  &782.501  &783.000  &782.501 \\
$m_\rho$       &763.0    &763.0    &770.0    &763.0  \\
$g_\sigma$       &9.6699   &10.217   &10.0289  &10.217 \\
$g_\omega$       &12.316   &12.868   &12.6139  &12.868 \\
$g_\rho(n_0)$         &14.1618  &8.948    &12.2413  &8.948  \\
$g_2/\rm{fm}^{-1}$    &11.9214  &10.431   &7.2325   &10.431  \\
$g_3$            &-31.6796 &-28.885  &0.6183   &-28.885 \\
$g_{\omega3}$   &2.6843   &0        &71.3075  &0   \\
$\Lambda_v$      &0.0475   &0        &0.0327   &0   \\
$a_\rho$               &0        &0.5835 &0        &0   \\
\hline\hline
\end{tabular}}
\end{table}

\section{Results and Discussions}
 \label{sec:results}
%
 \begin{figure*}[hptb]
    \centering
    \includegraphics[viewport=10 20 700 580, width=8 cm,clip]{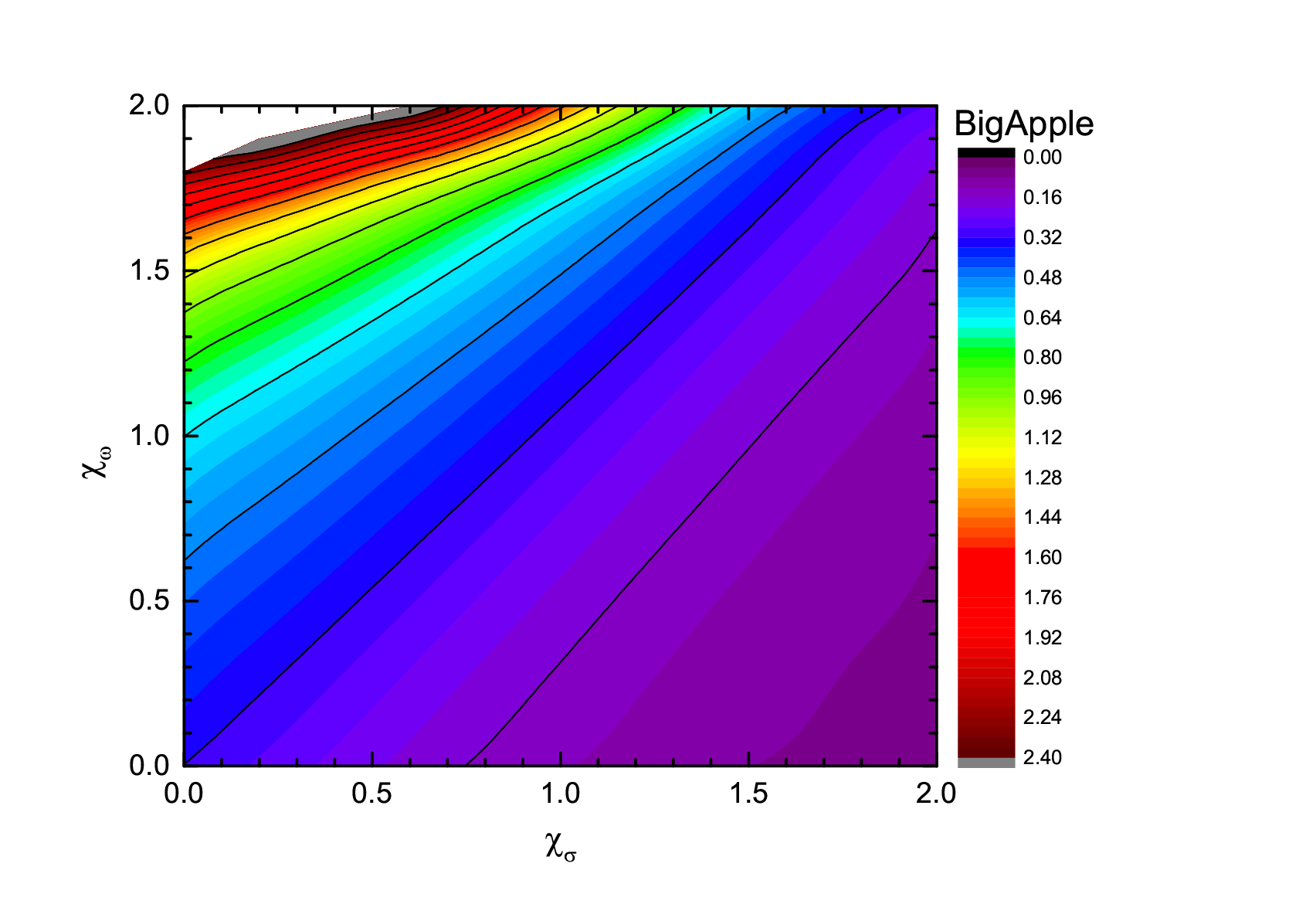}
    \includegraphics[viewport=10 20 700 580, width=8 cm,clip]{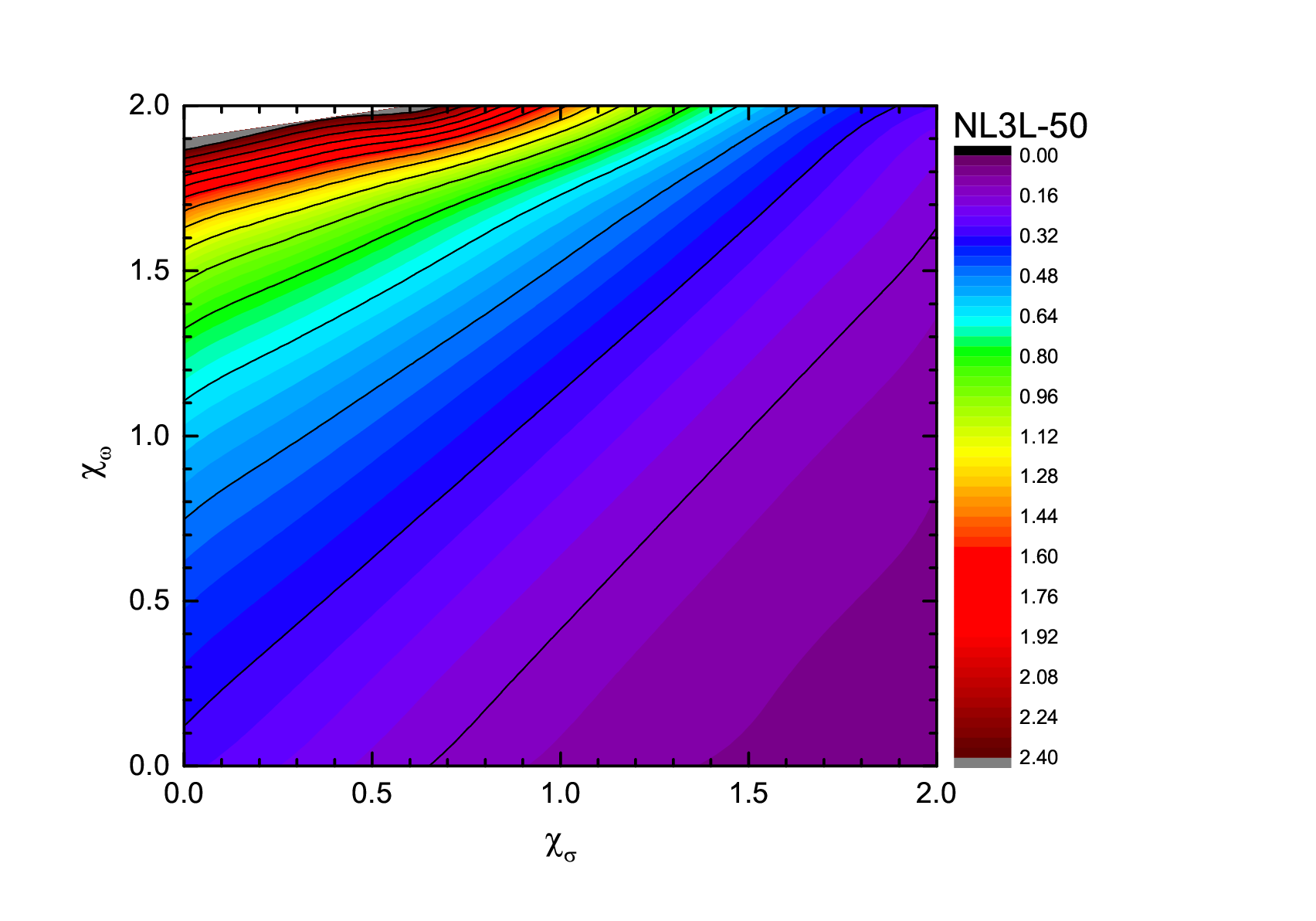}
    \includegraphics[viewport=10 20 700 580, width=8 cm,clip]{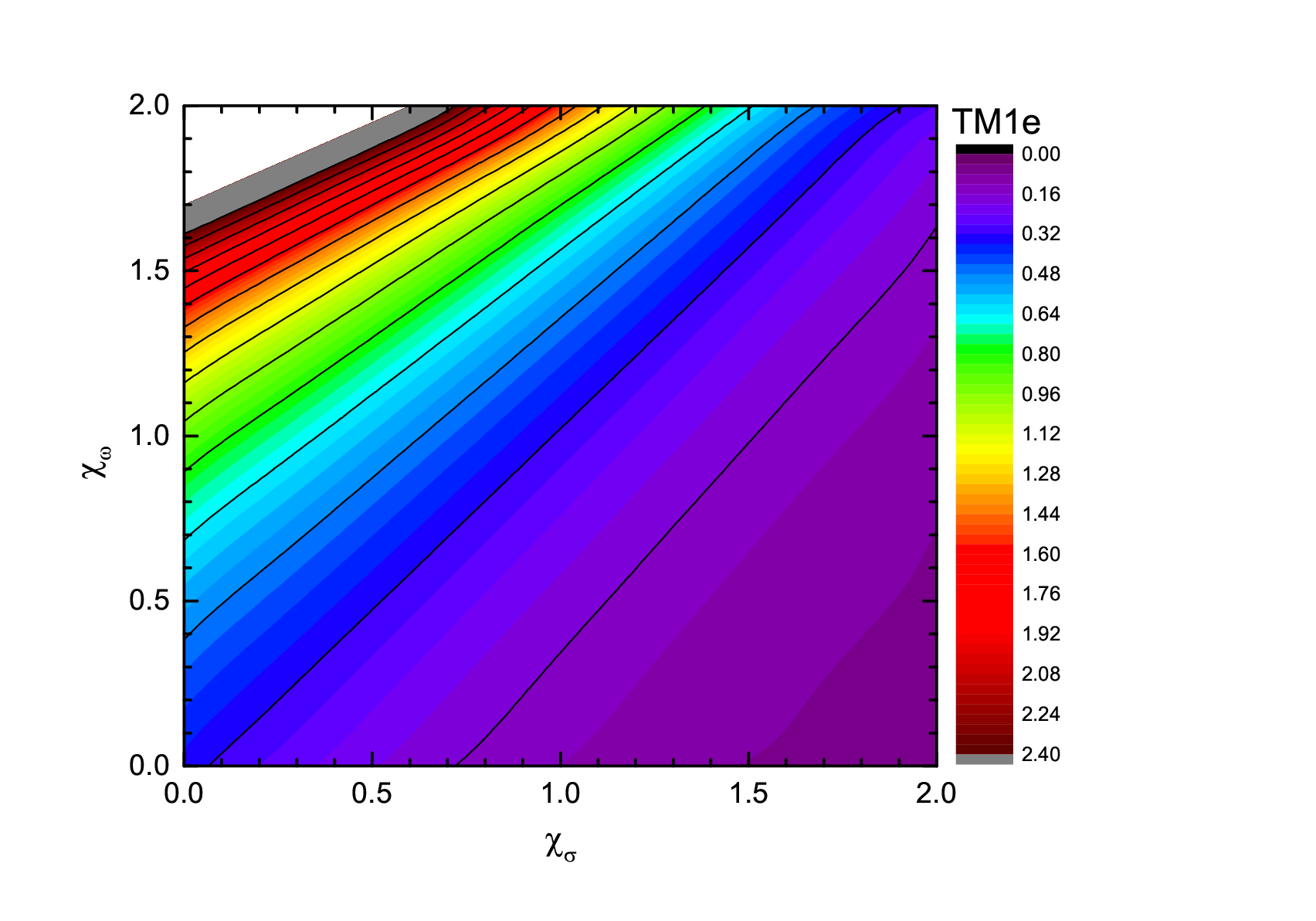}
    \includegraphics[viewport=10 20 700 580, width=8 cm,clip]{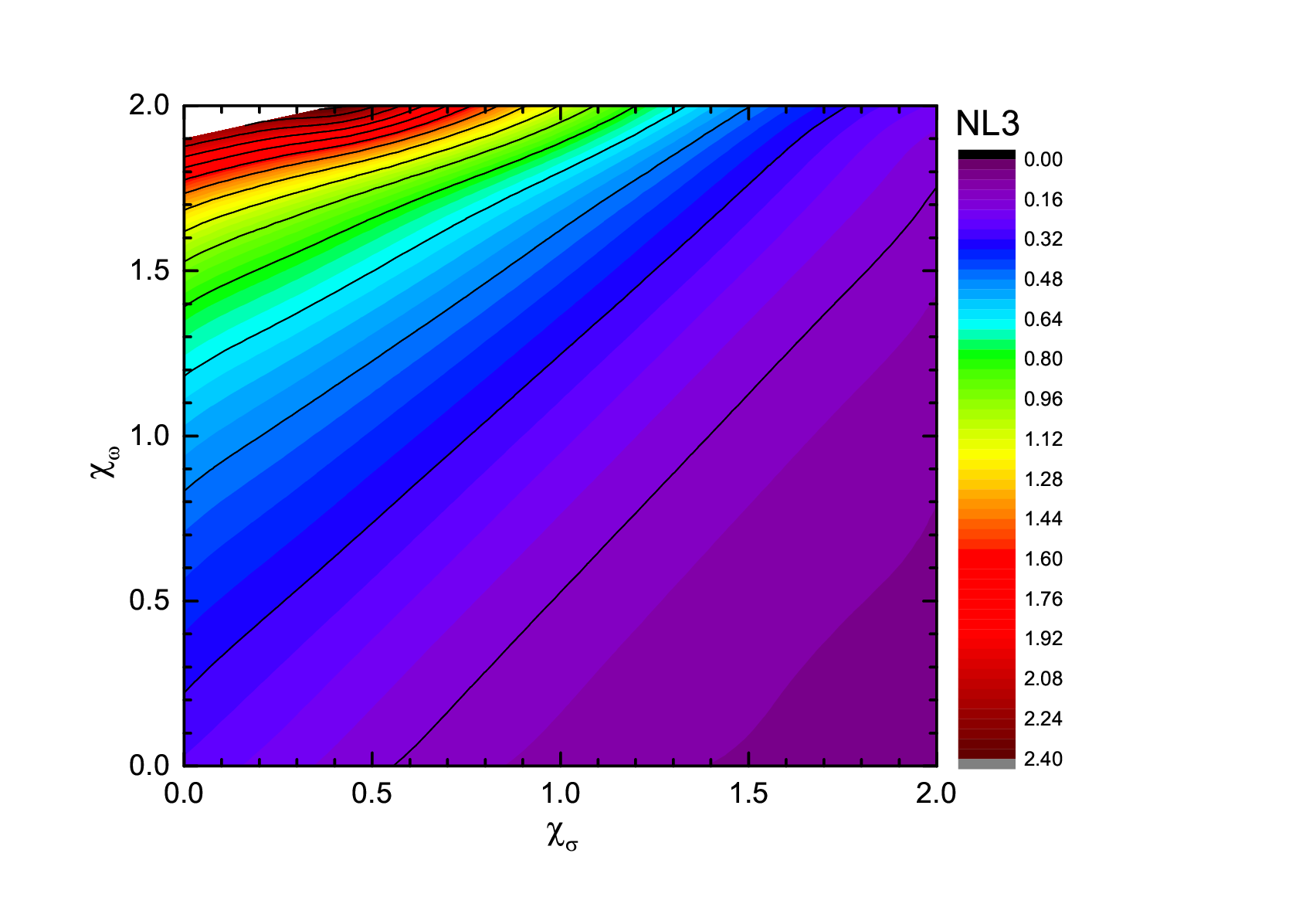}
    \caption{Onset density (in unit of $\rm{fm}^{-3}$) of  H particle in $\beta$-stable NS matter as a function of the dimensionless coupling ${\chi}_{\sigma H}$ and ${\chi}_{\omega H}$.}
    \label{fig:1-Happear}
\end{figure*}
 \begin{figure*}[hptb]
    \centering
    \includegraphics[viewport=10 10 580 580, width=8 cm,clip]{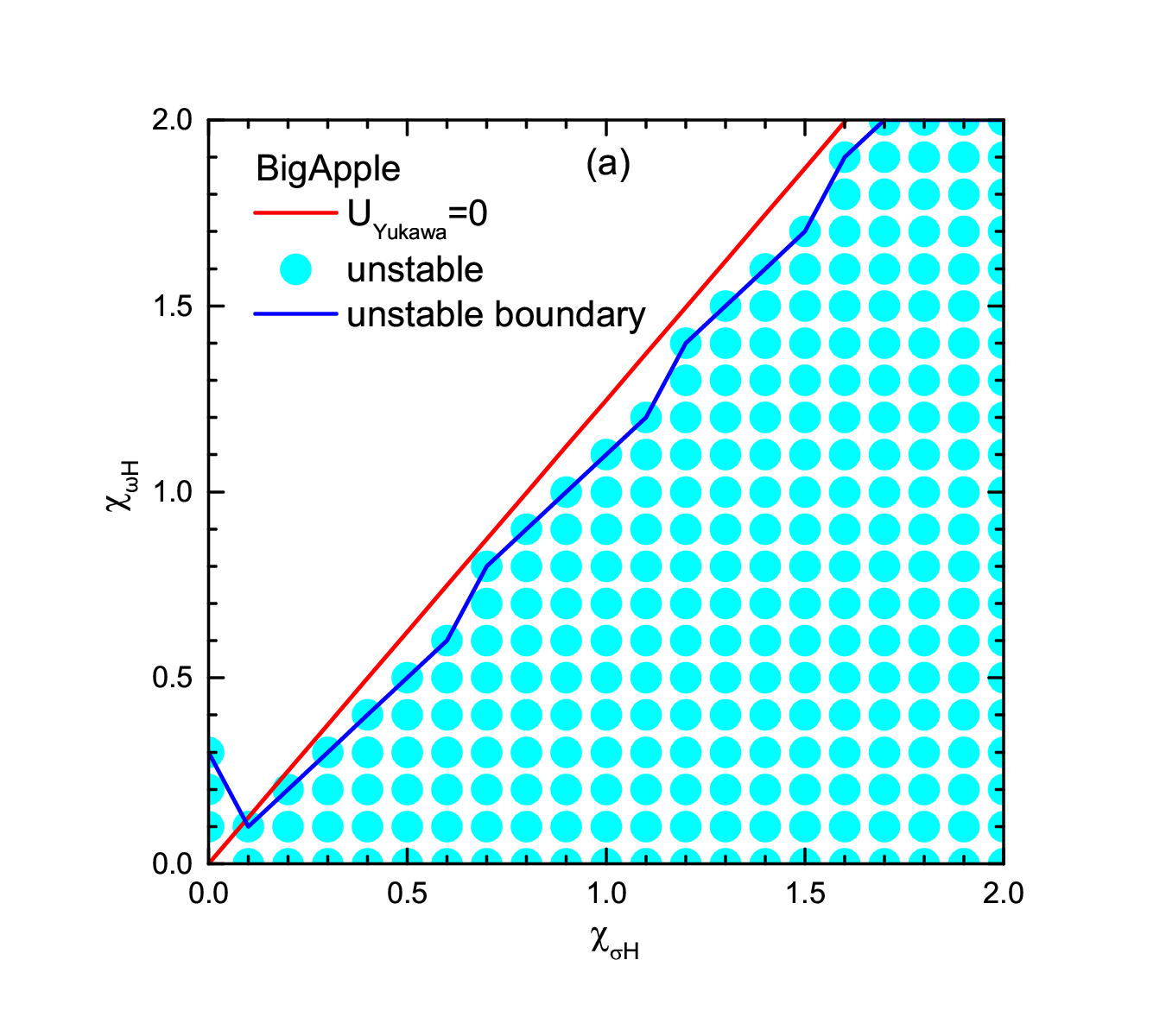}
    \includegraphics[viewport=10 10 580 580, width=8 cm,clip]{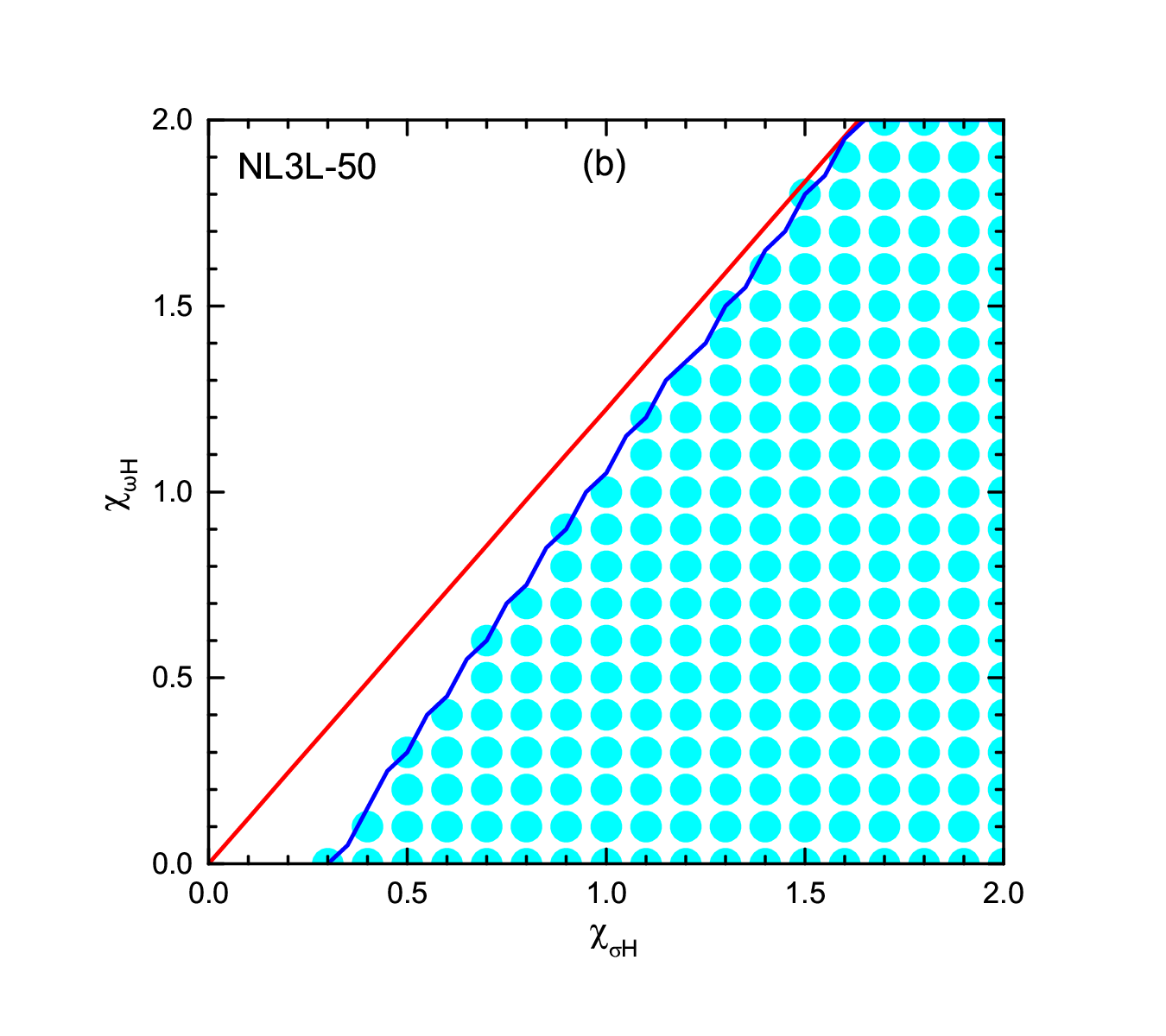}
    \includegraphics[viewport=10 10 580 580, width=8 cm,clip]{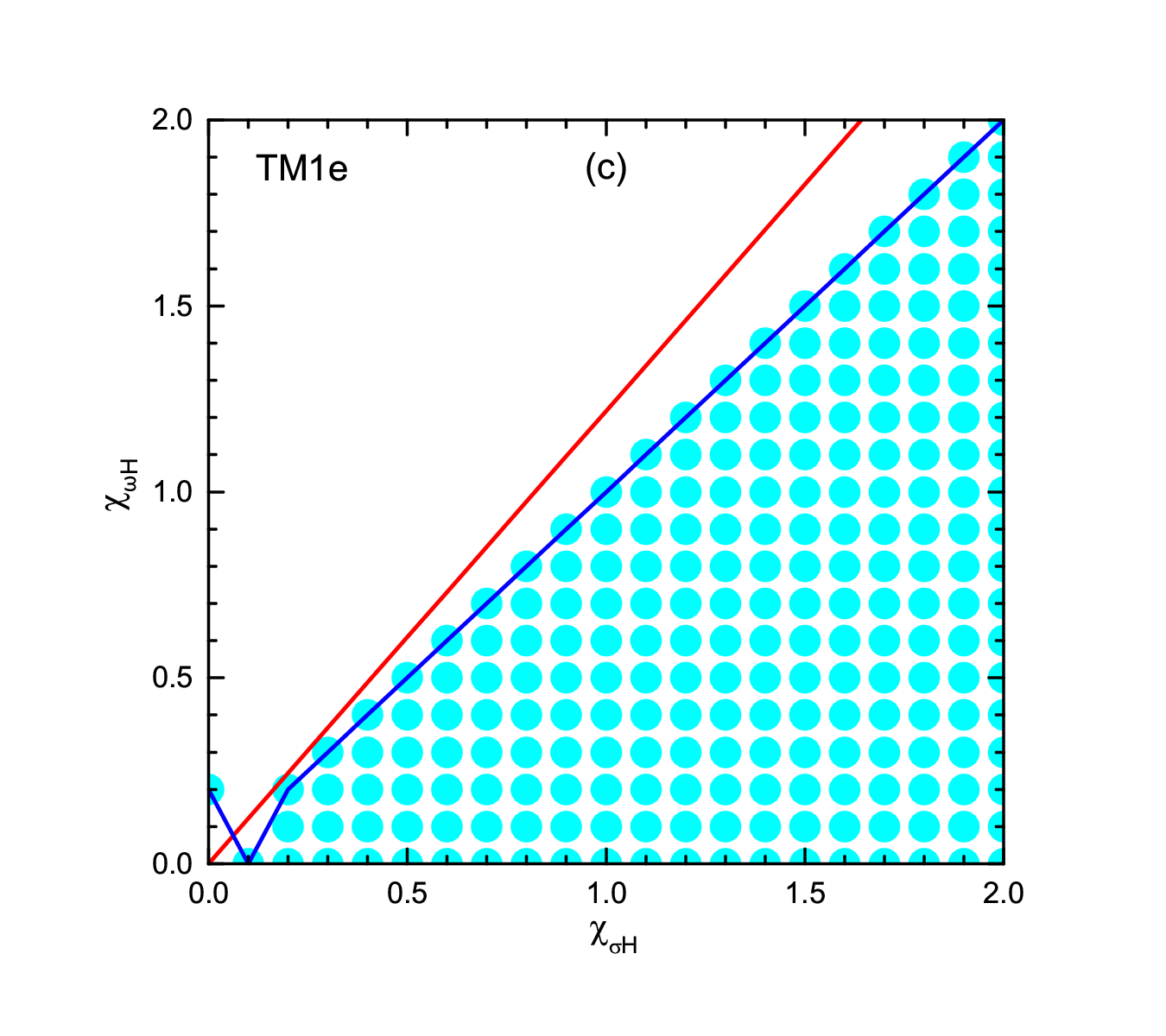}
    \includegraphics[viewport=10 10 580 580, width=8 cm,clip]{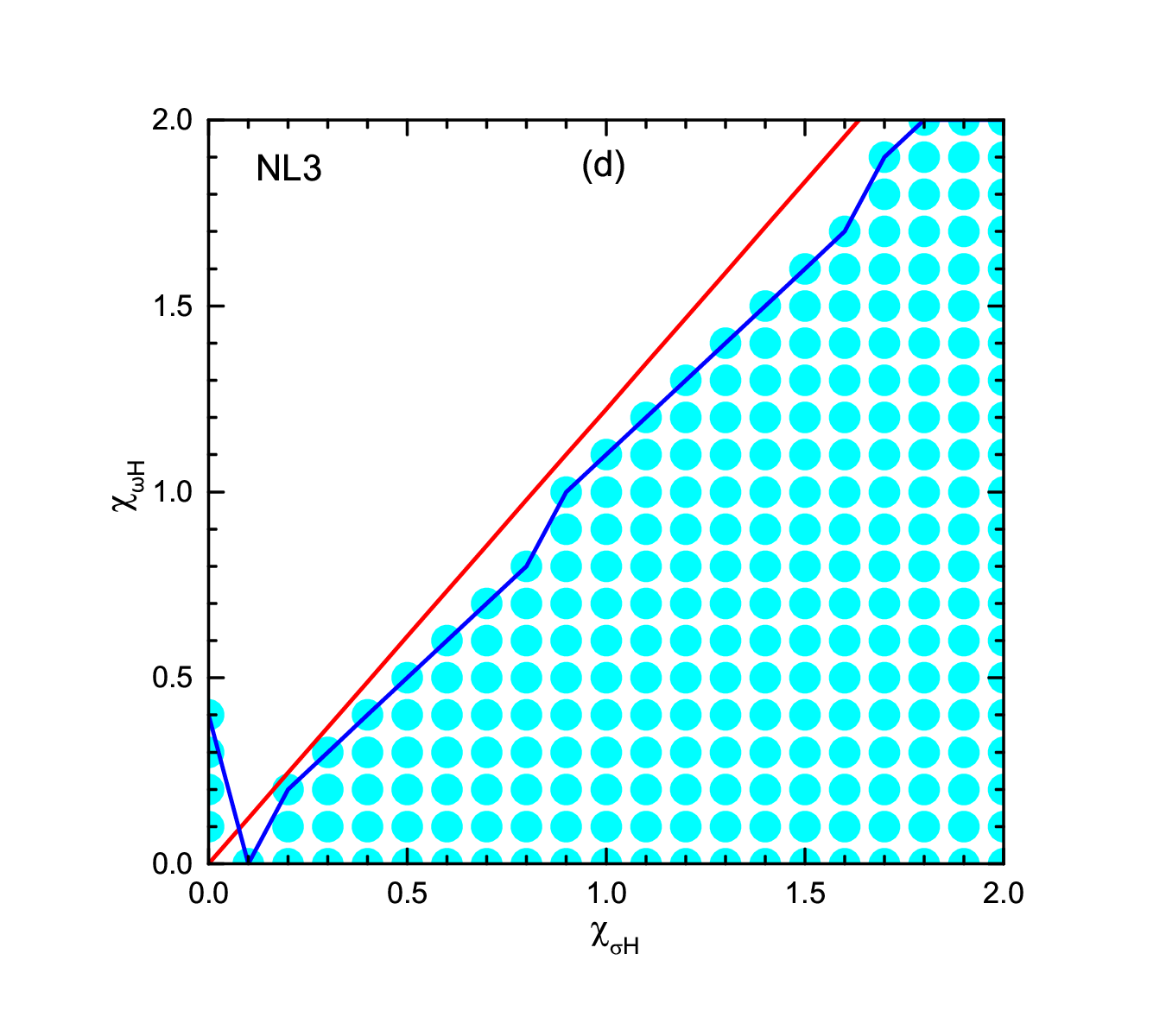}
    \caption{The solid blue line shows the area boundary of the zone where H particle's appearance leads to decreasing pressure.
    The red line corresponds to the critical boundary where Yukawa potential equals zero.}
    \label{fig:2-stable}
\end{figure*}
In Fig.~\ref{fig:1-Happear}, we show the onset density of the H particle as
a function of the dimensionless couplings $\chi_{\sigma H}$ and $\chi_{\omega H}$ within NS matter.
The nucleon component of the system is described using the BigApple, NL3L-50, TM1e, and NL3 parameter sets, respectively.
In the case of purely attractive interaction ($\chi_{\omega H}=0$),
the onset density of H particle decreases with $\chi_{\sigma H}$ increases.
Conversely, with a purely repulsive interaction ($\chi_{\sigma H}=0$),
the appearance of H particle is delayed to higher densities for larger values of $\chi_{\omega H}$.
When both $\chi_{\sigma H}$ and $\chi_{\omega H}$ are non-zero, the onset density could be connected to purely attractive case or purely repulsive case through the contour lines.
The NL3L-50, BigApple and NL3 parameter sets yield similar results for the onset density.
Comparing these with the results obtained using the relatively softer EOS TM1e, we find that the model dependence primarily stems from the stiffness of the EOS but have little dependence on the symmetry energy.
With stiffer EOS, the onset density of H particle is smaller.

The stability of NS containing H particle is displayed in Fig.~\ref{fig:2-stable}.
The light blue region, labeled with the corresponding coupling strengths $\chi_{\sigma H}$ and $\chi_{\omega H}$ (step size $0.1$), indicates a decrease in pressure upon the appearance of H particle.
In other words, the presence of the H particle with the couplings of the light blue region
leads to an instability in the many-body system which may result in collapse of NS.
The blue solid line represents the boundary of the unstable region.
The Yukawa potential can be defined according to Ref.~\cite{Faessler1997},
which requires
\begin{eqnarray}
\frac{g_{\sigma H}^2}{m_{\sigma}^2}<\frac{g_{\omega H}^2}{m_{\omega}^2}.
\end{eqnarray}
The boundary where H particle does not contribute to the pressure, namely
$U_{\rm{Yukawa}}=\frac{g_{\sigma H}^2}{m_{\sigma}^2}-\frac{g_{\omega H}^2}{m_{\omega}^2}=0$, is presented by the red solid line.
There is a noticeable gap between the red and blue lines,
indicating that despite the negative contribution from H particle,
the strong repulsive interaction at high density and the Fermi degeneracy pressure from nucleons and leptons are able to sustain a stable system.
However, this stability does not persist at higher densities, even with $U_{\rm{Yukawa}}>0$ (located in the upper left of the red line),
because the pressure provided by H particle is too small compared to other particles, and its fraction increases with density.
On one hand, the fractions of nucleons and leptons decrease,
which lead to a reduction of their contributions to the pressure.
On the other hand, H particle contributes only a minimal amount of pressure,
even if it has a positive value,
but it cannot balance the reduction from other particles.
There is a extreme case that H particle could exhibit a total repulsive interaction similar in magnitude to nucleons,
but under such conditions, the appearance of H particle would be significantly delayed beyond the NS central density.
As a result, the pressure of NS matter containing H particles destines to decrease with increasing density, sooner or later.

 \begin{figure*}[hptb]
    \centering
    \includegraphics[viewport=10 10 580 430, width=8 cm,clip]{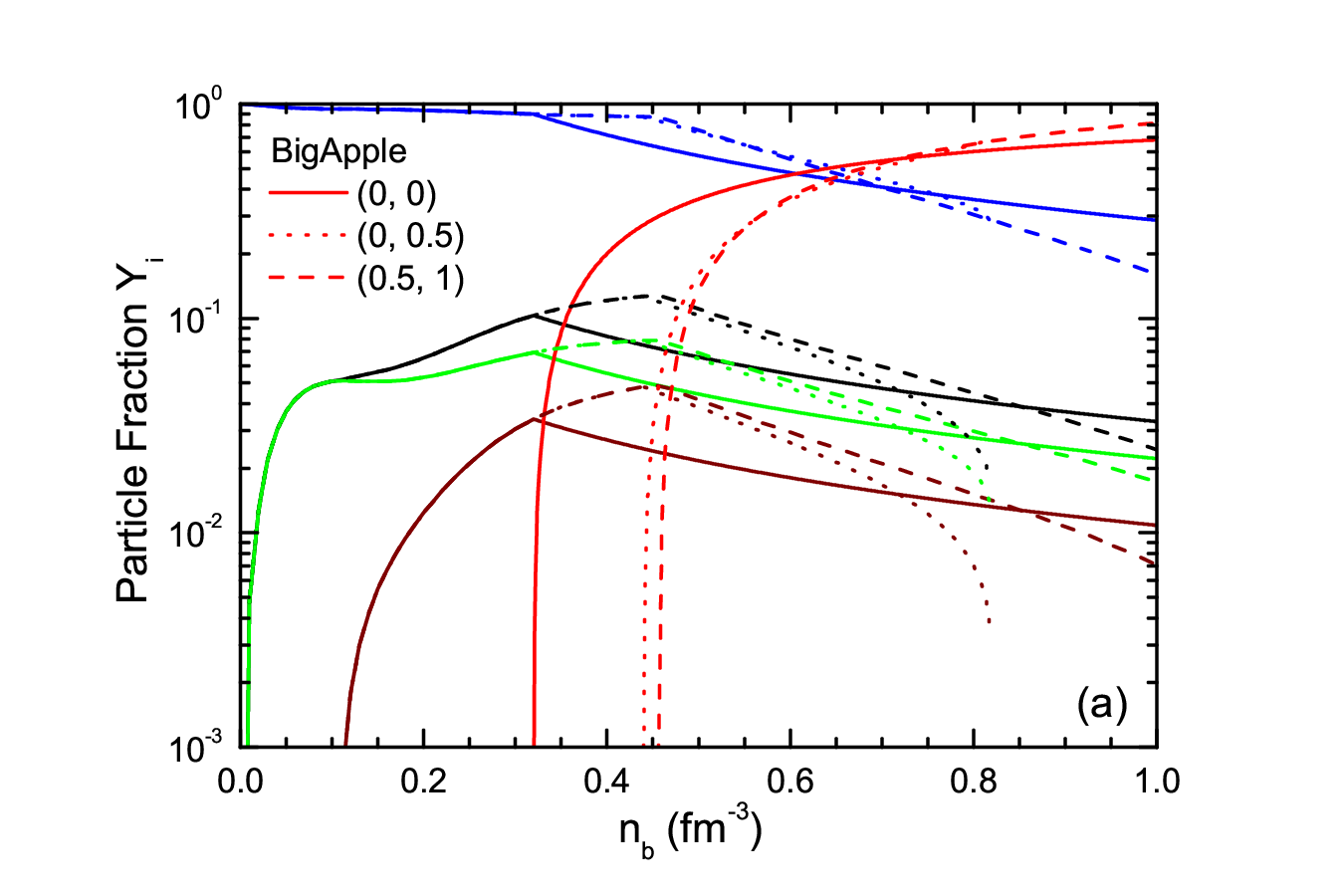}
    \includegraphics[viewport=10 10 580 430, width=8 cm,clip]{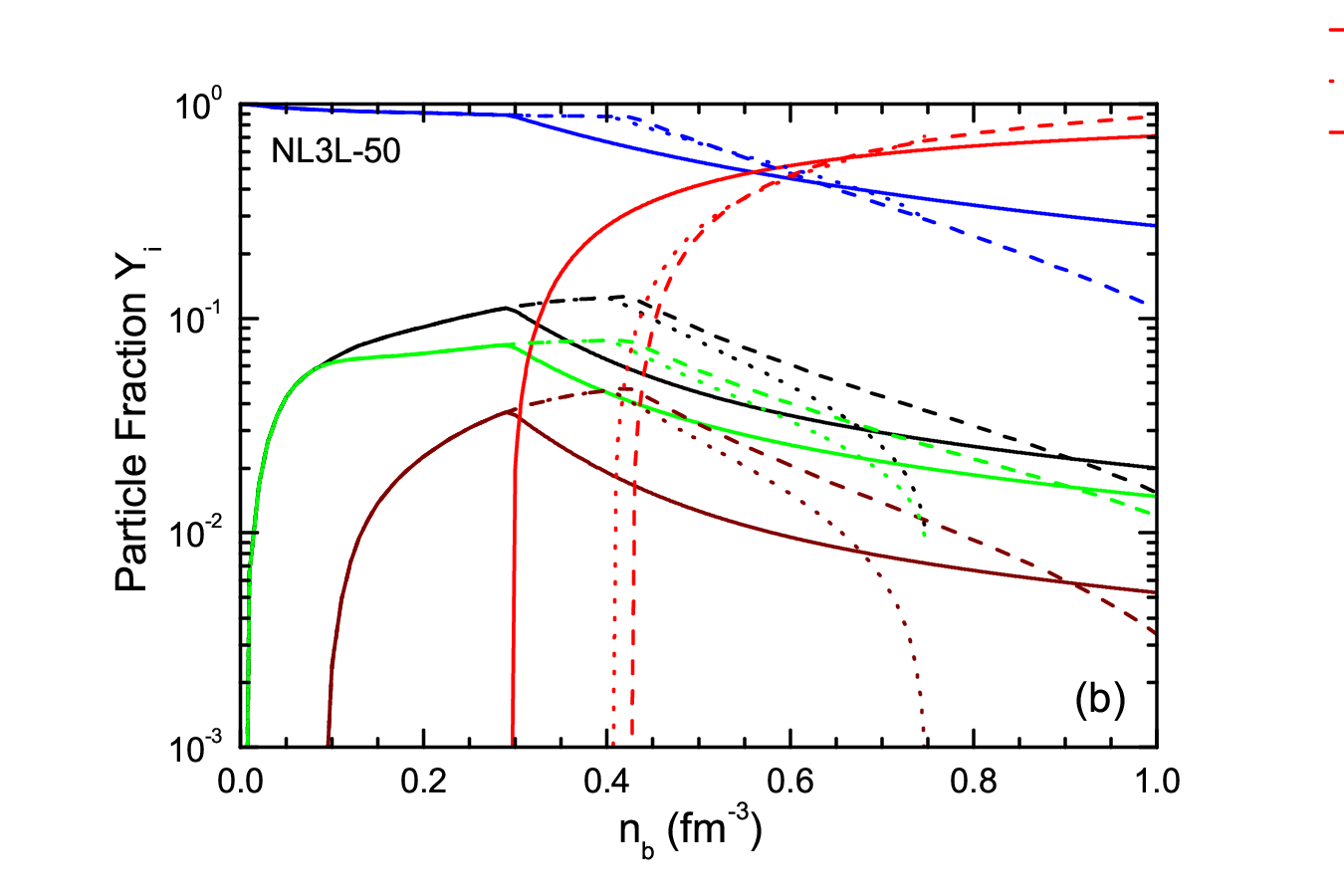}
    \includegraphics[viewport=10 10 580 430, width=8 cm,clip]{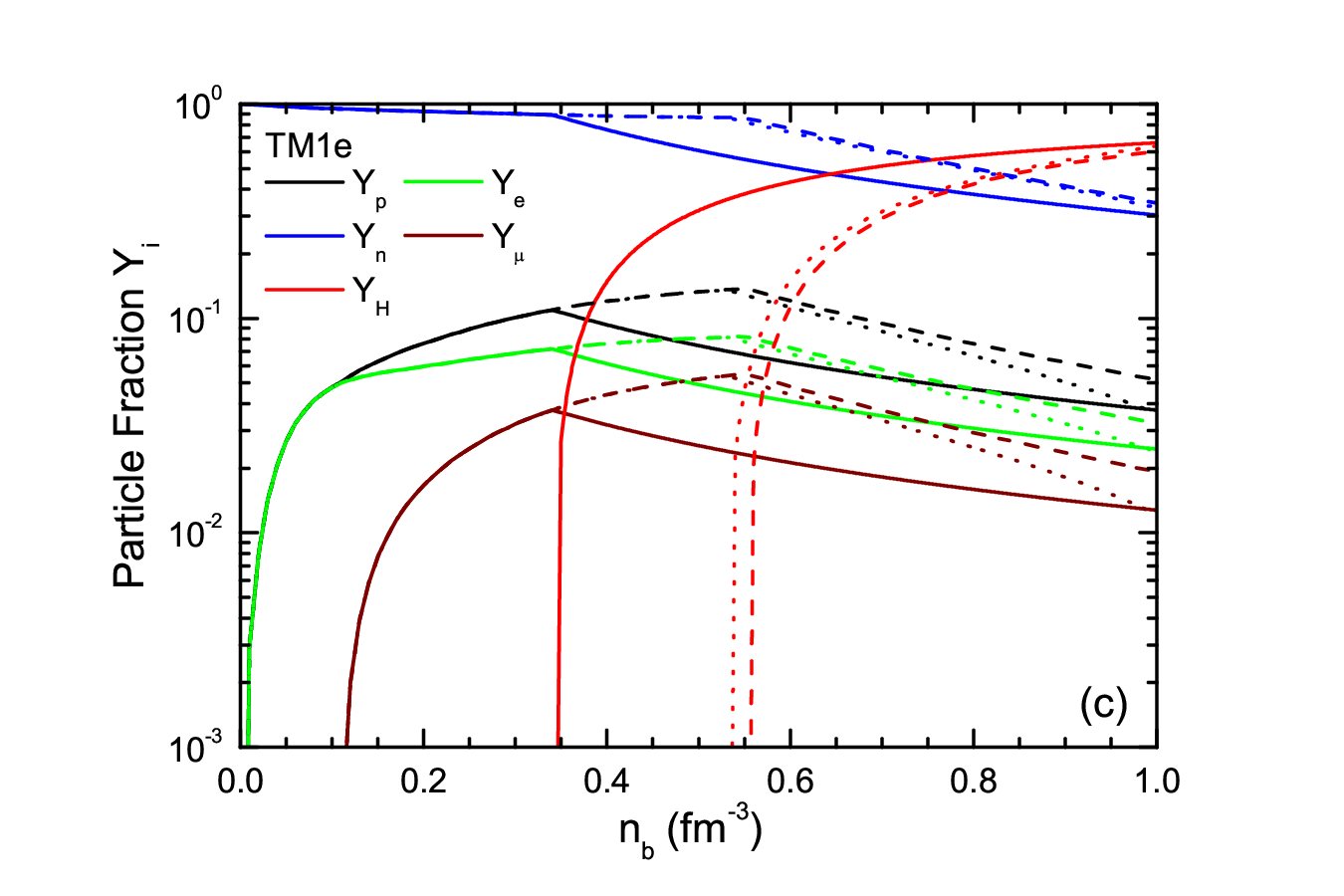}
    \includegraphics[viewport=10 10 580 430, width=8 cm,clip]{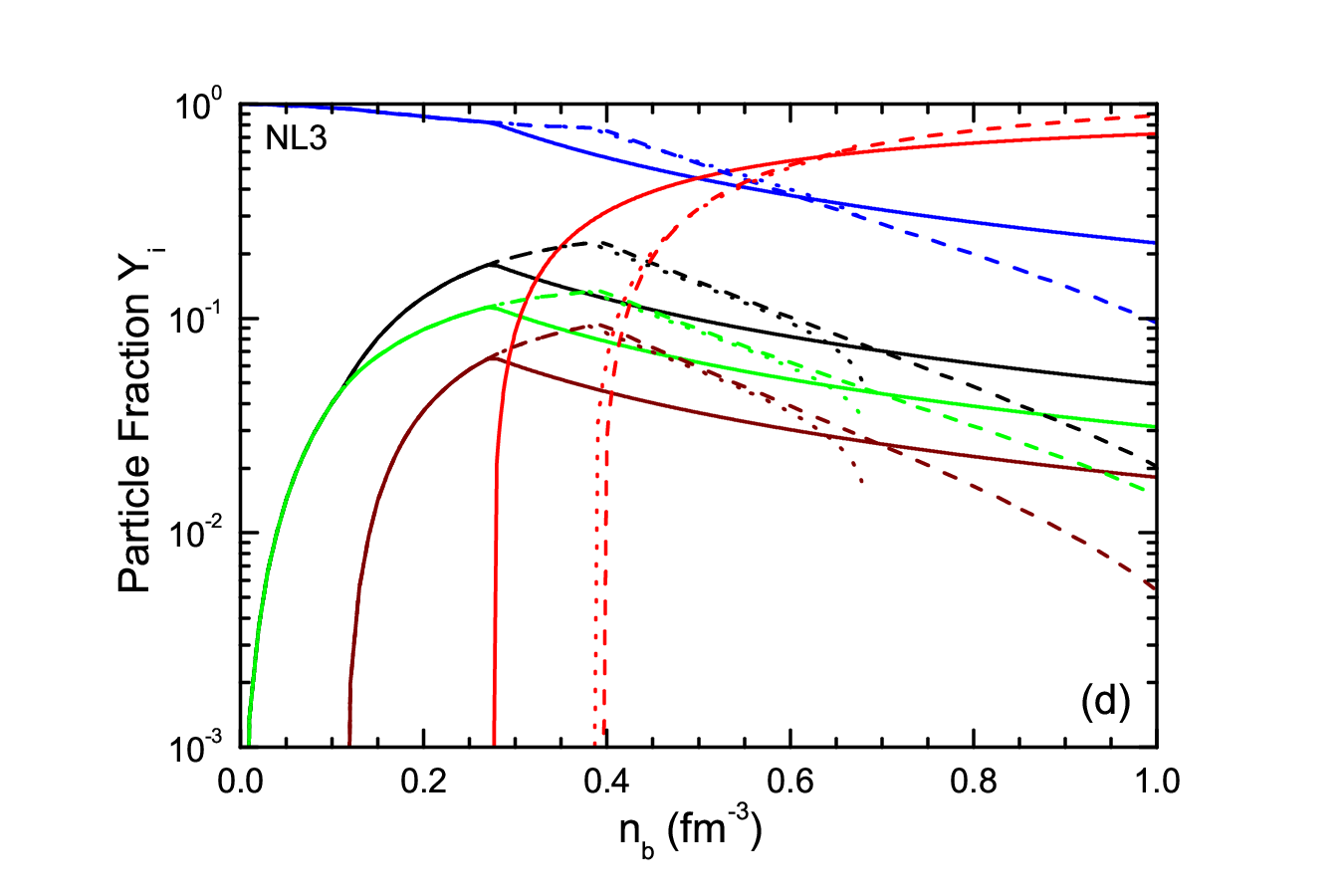}
    \caption{Particle fractions of protons, neutrons, H particles, electrons, and muons ($Y_p, Y_n, Y_H, Y_e, Y_\mu$) as functions of the baryon number density $n_b$ in different models.
    The three cases of H particle couplings used, (${\chi}_{\sigma H}$, ${\chi}_{\omega H}$)=(0, 0), (0, 0.5), (0.5, 1),
    correspond to free H particle, purely repulsive interaction and both attractive and repulsive interaction.}
    \label{fig:3-fraction}
\end{figure*}
 \begin{figure*}[hptb]
    \centering
    \includegraphics[viewport=10 10 580 430, width=8 cm,clip]{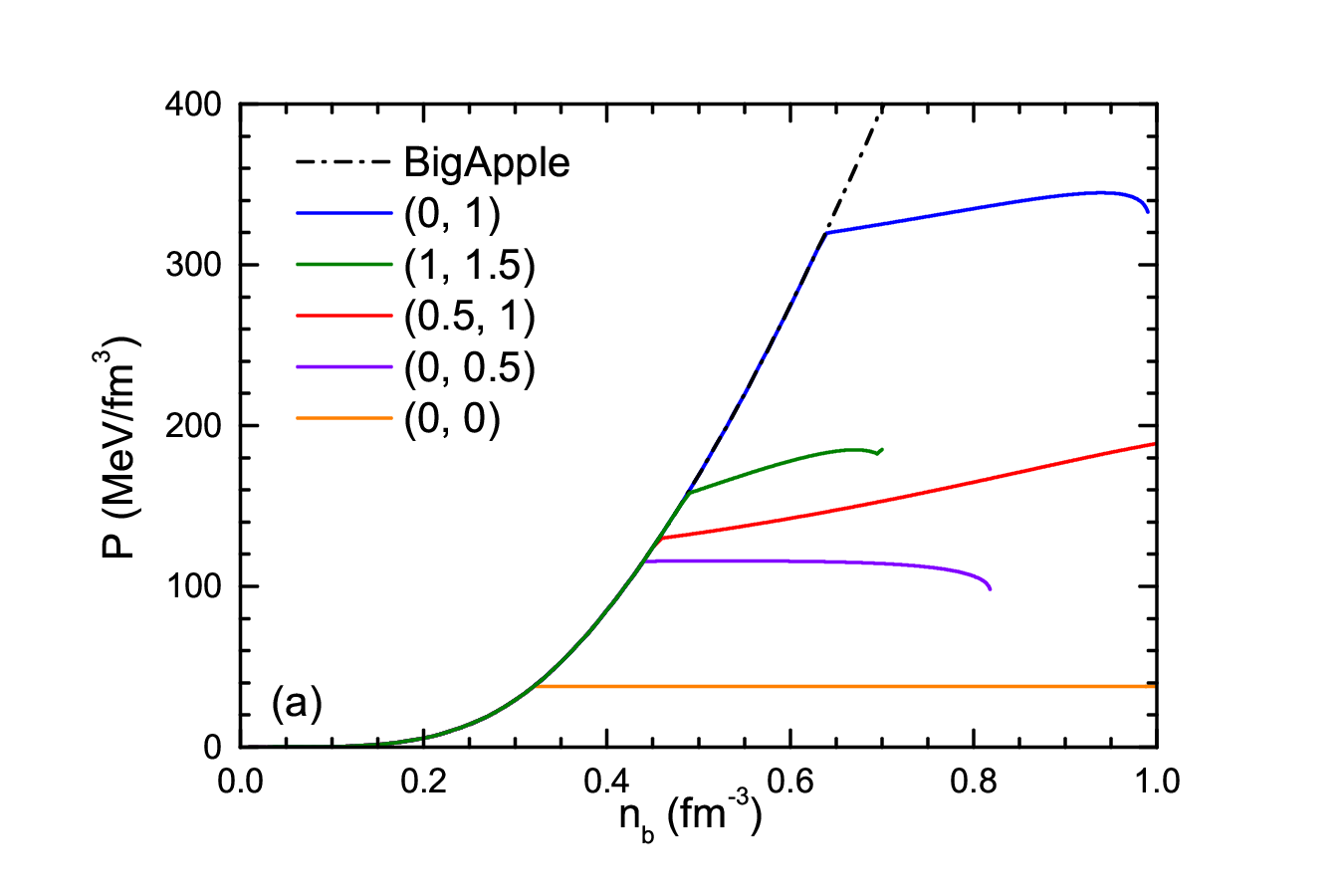}
    \includegraphics[viewport=10 10 580 430, width=8 cm,clip]{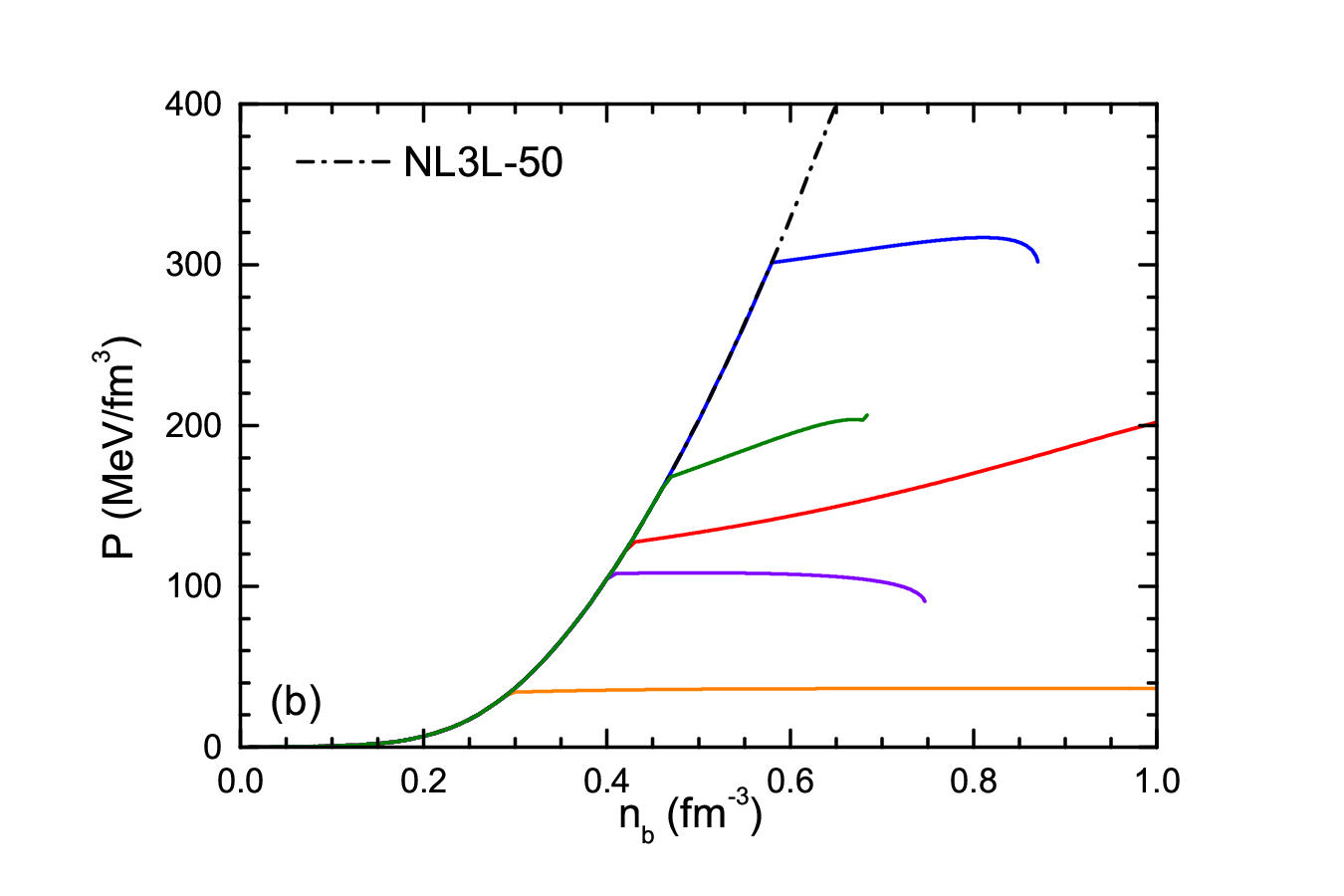}
    \includegraphics[viewport=10 10 580 430, width=8 cm,clip]{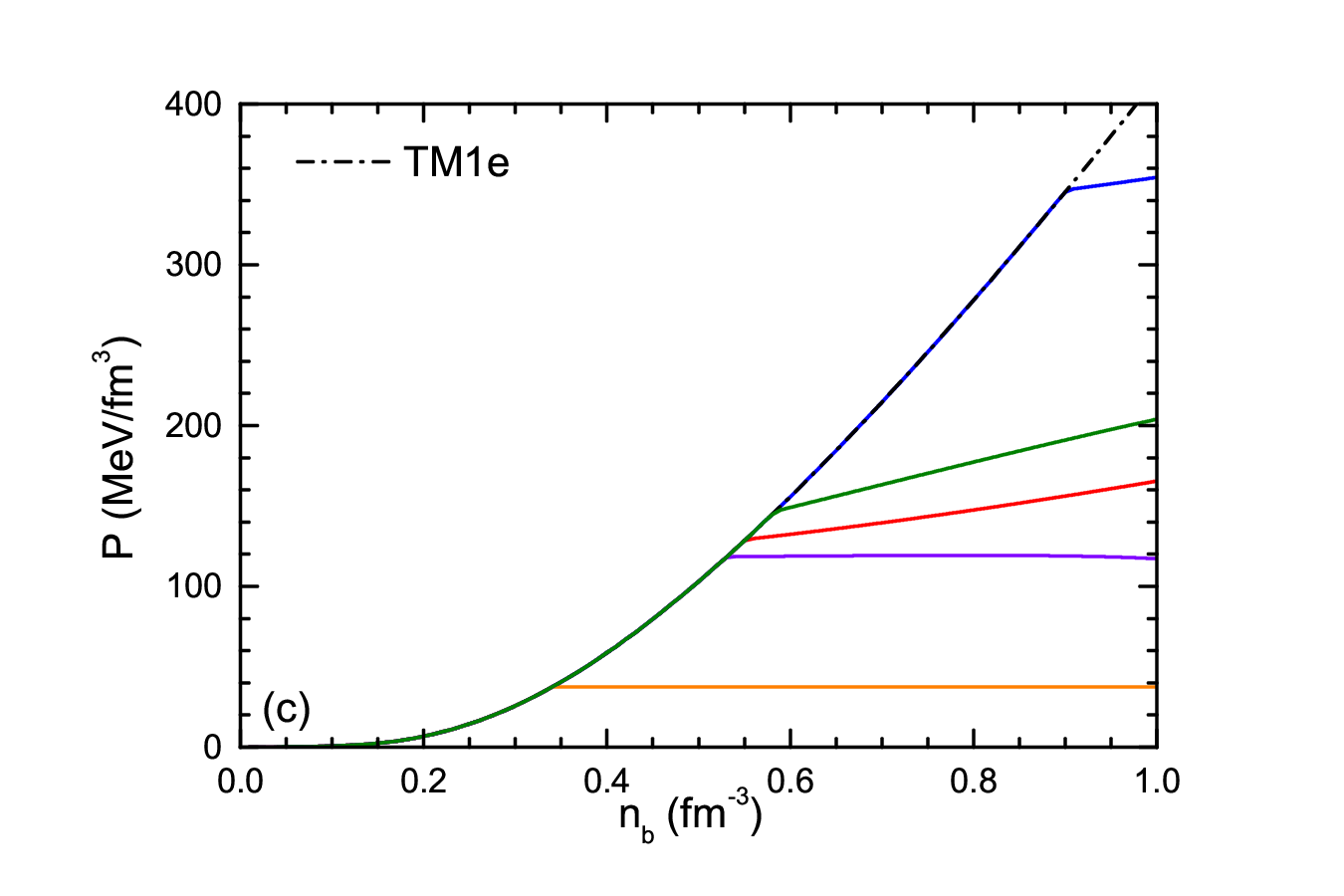}
    \includegraphics[viewport=10 10 580 430, width=8 cm,clip]{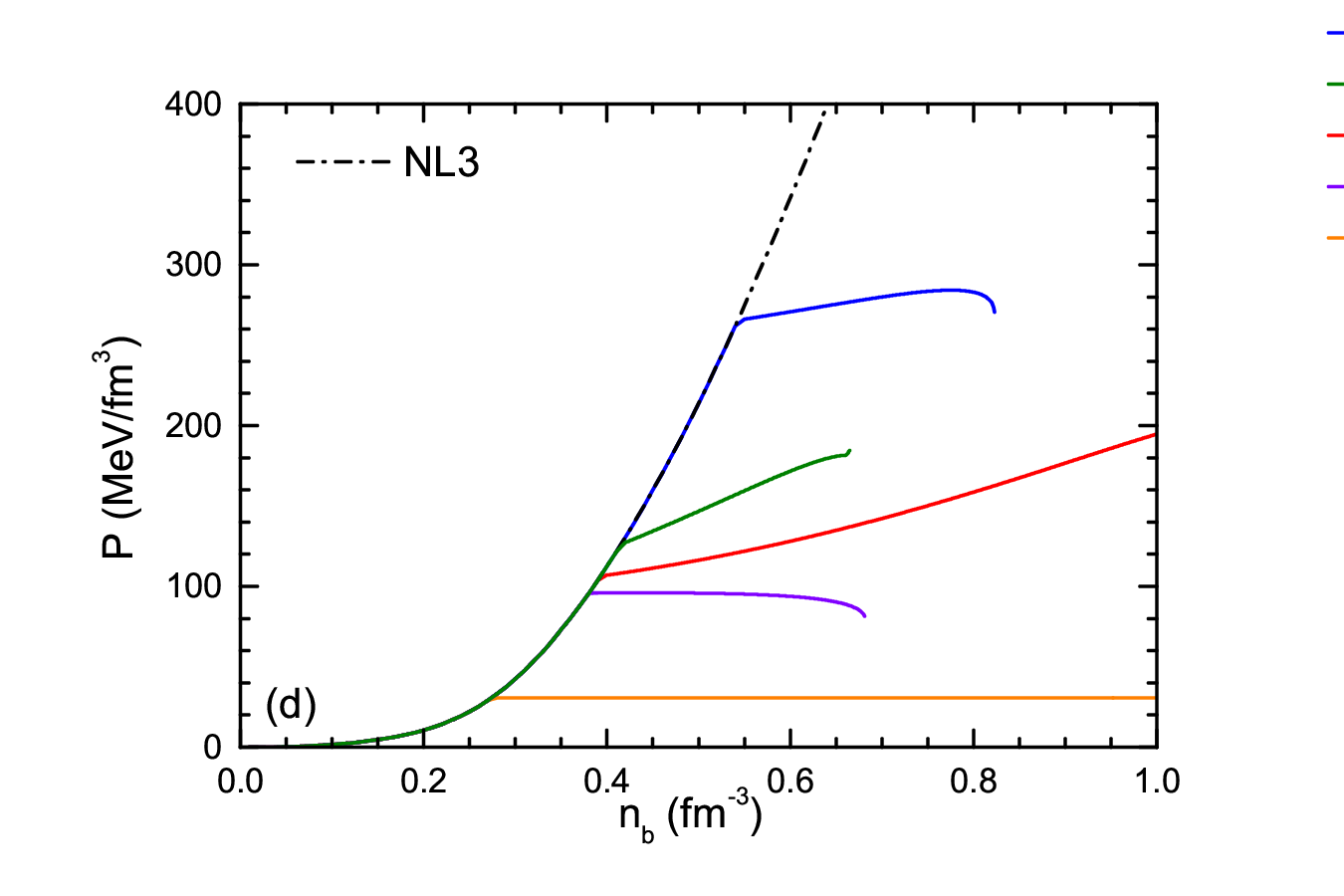}
    \caption{Pressure $P$ as a function of $n_b$ with different couplings.
    The results using couplings (${\chi}_{\sigma H}$, ${\chi}_{\omega H}$)=(0, 0), (0, 0.5), (0.5, 1), (1, 1.5), (0, 1) are shown.
    H particle admixed pressures split from those of pure hadronic matter when the H particles appear.}
    \label{fig:4-nbp}
\end{figure*}

The particle fractions as a function of the baryon number density $n_b$ are displayed in Fig.~\ref{fig:3-fraction}.
Upon the appearance of H particle, the fractions of all other types of particles, such as nucleons and leptons, immediately decrease.
Our results are different from the behavior of another hexaquark $d^*(2380)$ shown in Ref.~\cite{Mantziris2020}, where the appearance of $d^*(2380)$ leads to an increasing fraction of leptons.
This is because $d^*(2380)$ is positively charged,
therefore, its appearance requires an increasing lepton fraction in order to
maintain the charge neutrality.
Consequently, the lepton degeneracy pressure contributes positively to the total pressure,
resulting in a slower decrease in the pressure of NS matter due to the appearance of $d^*(2380)$,
compared to that containing H particle.
Since the H particle is electrically neutral, its presence results in a higher proportion of H particles as the density increases,
which leads to a decrease in the fractions of neutrons and protons.
The fraction of leptons decreases synchronously with the fraction of protons.
With increasing density, the process of neutronization in NS can be considered as a decrease in the electron fraction.
This situation is also applicable to deconfinement phase transitions.
In this study, the appearance of the H particle promotes the reduction of electron number,
thereby squeezing out the proportion of nucleons.
The purely repulsive interaction $(\chi_{\sigma H}, \chi_{\omega H})=(0, 0.5)$ of H particles causes the fractions of other particles to decrease more rapidly
(The curves depicting $(\chi_{\sigma H}, \chi_{\omega H})=(0, 0.5)$ terminate around 0.7$-$0.8 fm$^{-3}$ in panels a, b, and d due to the occurrence of collapse,
see details in Fig.~\ref{fig:4-nbp}).
Including attractive interaction $(\chi_{\sigma H}, \chi_{\omega H})=(0.5, 1)$ increases the fraction of H particles.
The influence of different couplings on the onset of the H particle can be found in Fig.~\ref{fig:1-Happear}.
In general, a total repulsive interaction delays the appearance of the H particle.

In Fig.~\ref{fig:4-nbp}, we present the pressures obtained using different RMF/RMFL models, considering various interactions of H particles.
The interaction of H particles is assumed to be either purely repulsive with couplings $(\chi_{\sigma H}, \chi_{\omega H})=(0, 0.5), (0, 1)$ or totally repulsive with couplings $(\chi_{\sigma H}, \chi_{\omega H})=(0.5, 1), (1, 1.5)$.
For comparison, the non-interacting case $(\chi_{\sigma H}, \chi_{\omega H})=(0, 0)$ is also included.
The condition of totally attractive interactions is not depicted here,
as the immediate reduction in pressure does not support the stable existence of H particles (except part of NL3L-50 results in Fig.~\ref{fig:2-stable}).
The onset of H particle can also be observed in Fig.~\ref{fig:1-Happear}.
Clearly, the system becomes unstable at high density under the conditions $(\chi_{\sigma H}, \chi_{\omega H})=(0, 0.5), (1, 1.5), (0, 1)$.
Other cases in Fig.~\ref{fig:4-nbp} like $(\chi_{\sigma H}, \chi_{\omega H})=(0.5, 1)$ will also decrease at very high density.
If there is an increase in pressure following the appearance of the H particle,
it could survive in NS matter at least for a range of density.

\begin{figure*}[hptb]
    \centering
    \includegraphics[viewport=10 20 580 570, width=8 cm,clip]{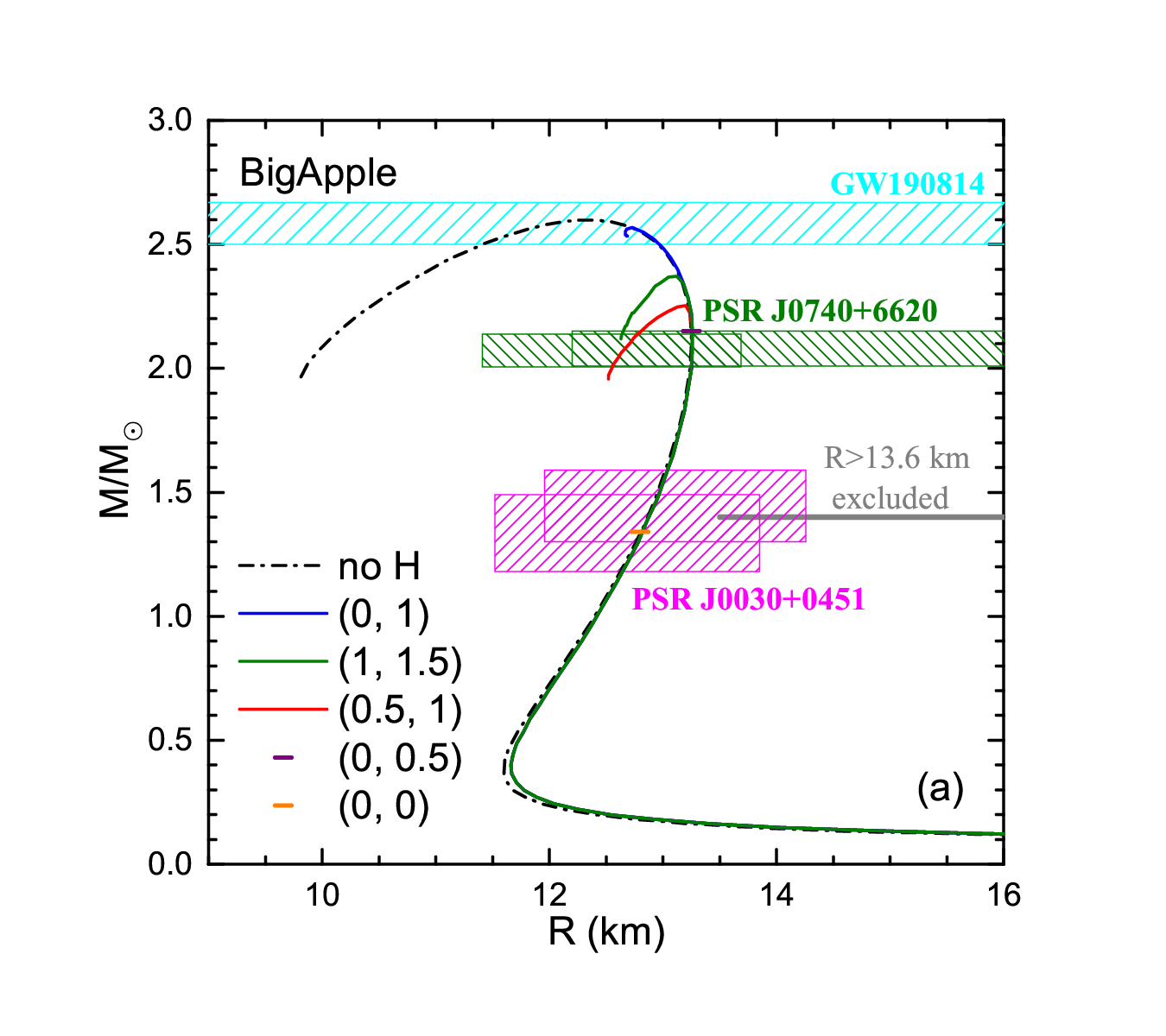}
    \includegraphics[bb=10 20 580 570, width=8 cm,clip]{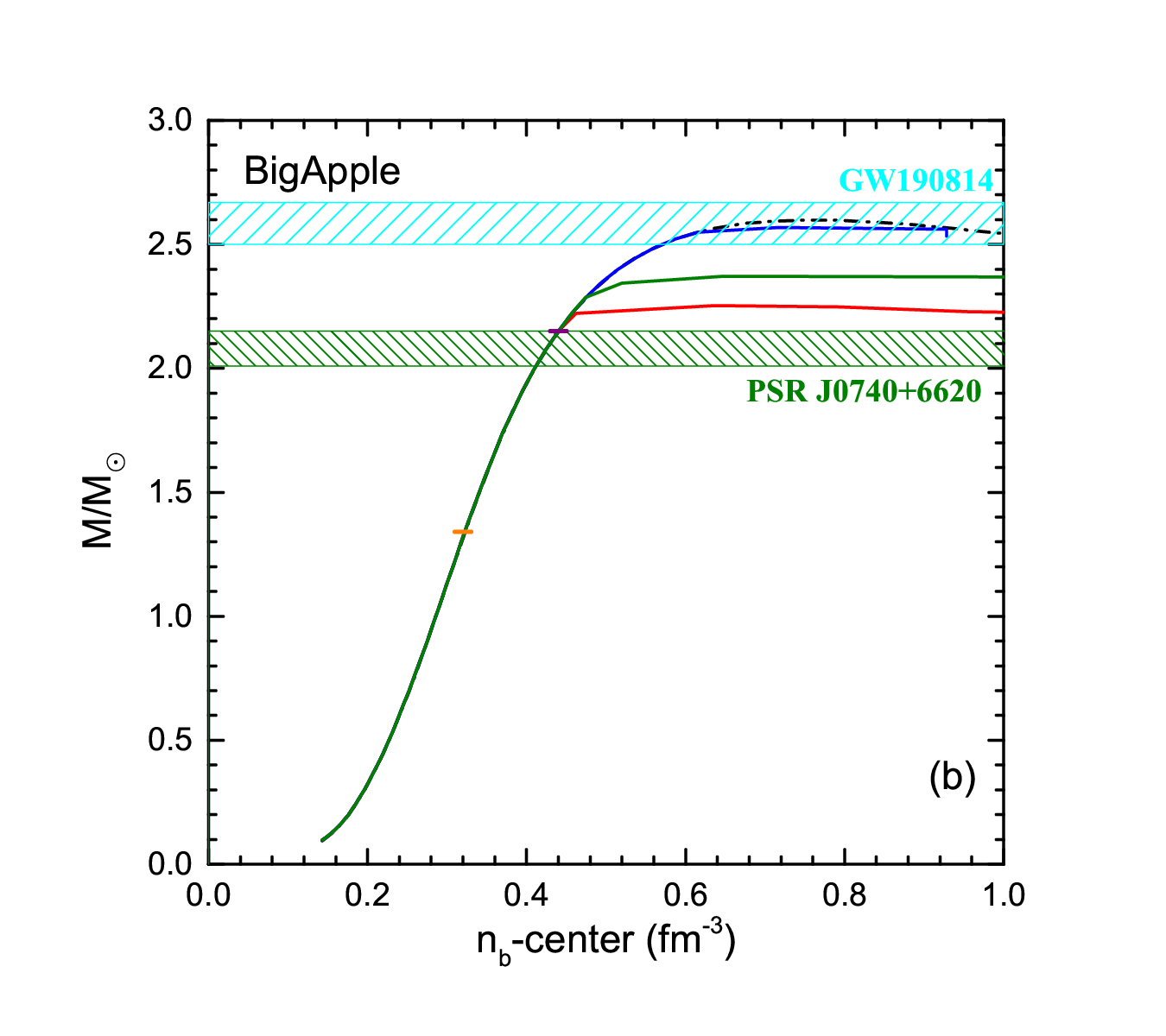} \\
    \includegraphics[viewport=10 20 580 570, width=8 cm,clip]{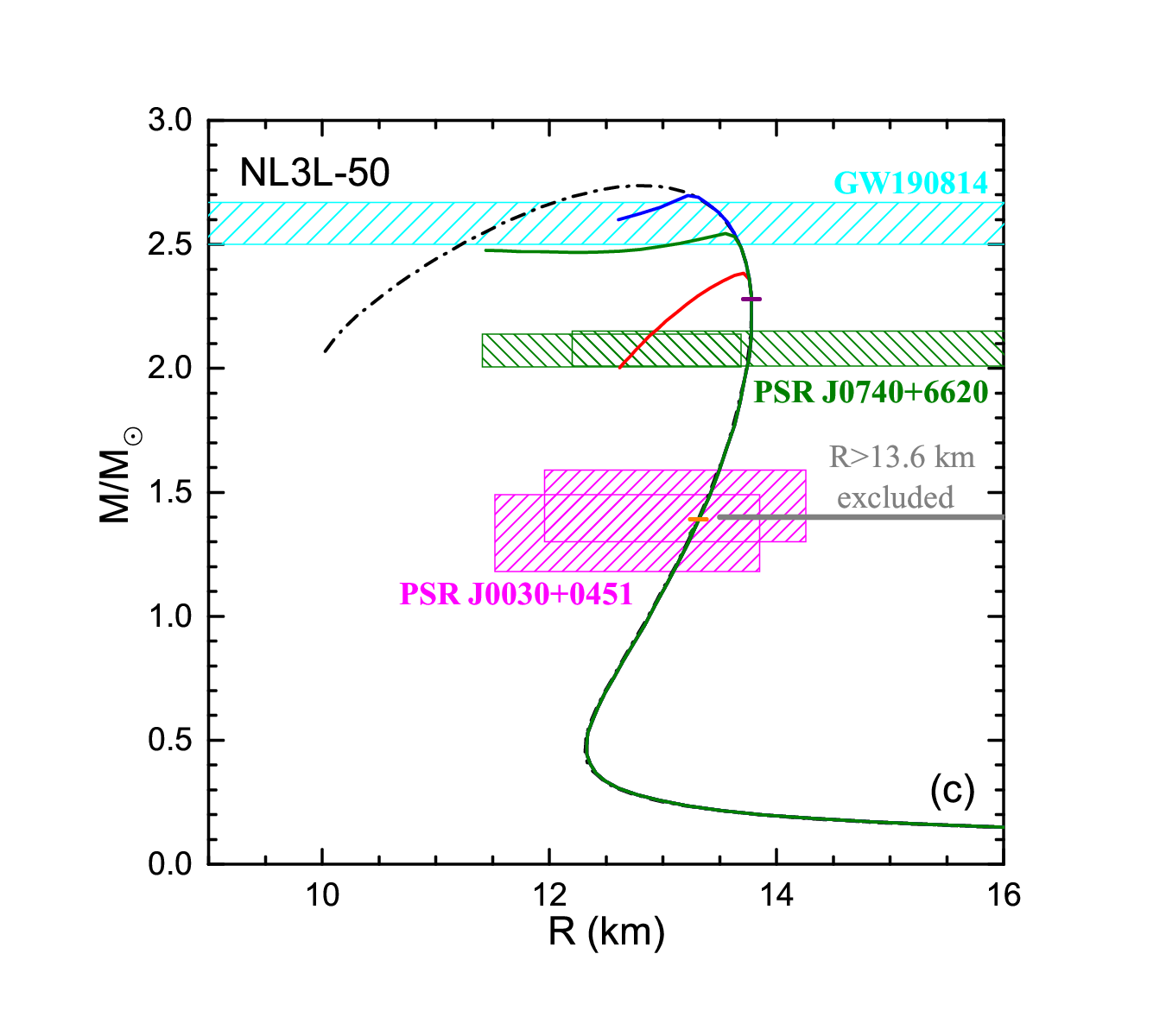}
    \includegraphics[bb=10 20 580 570, width=8 cm,clip]{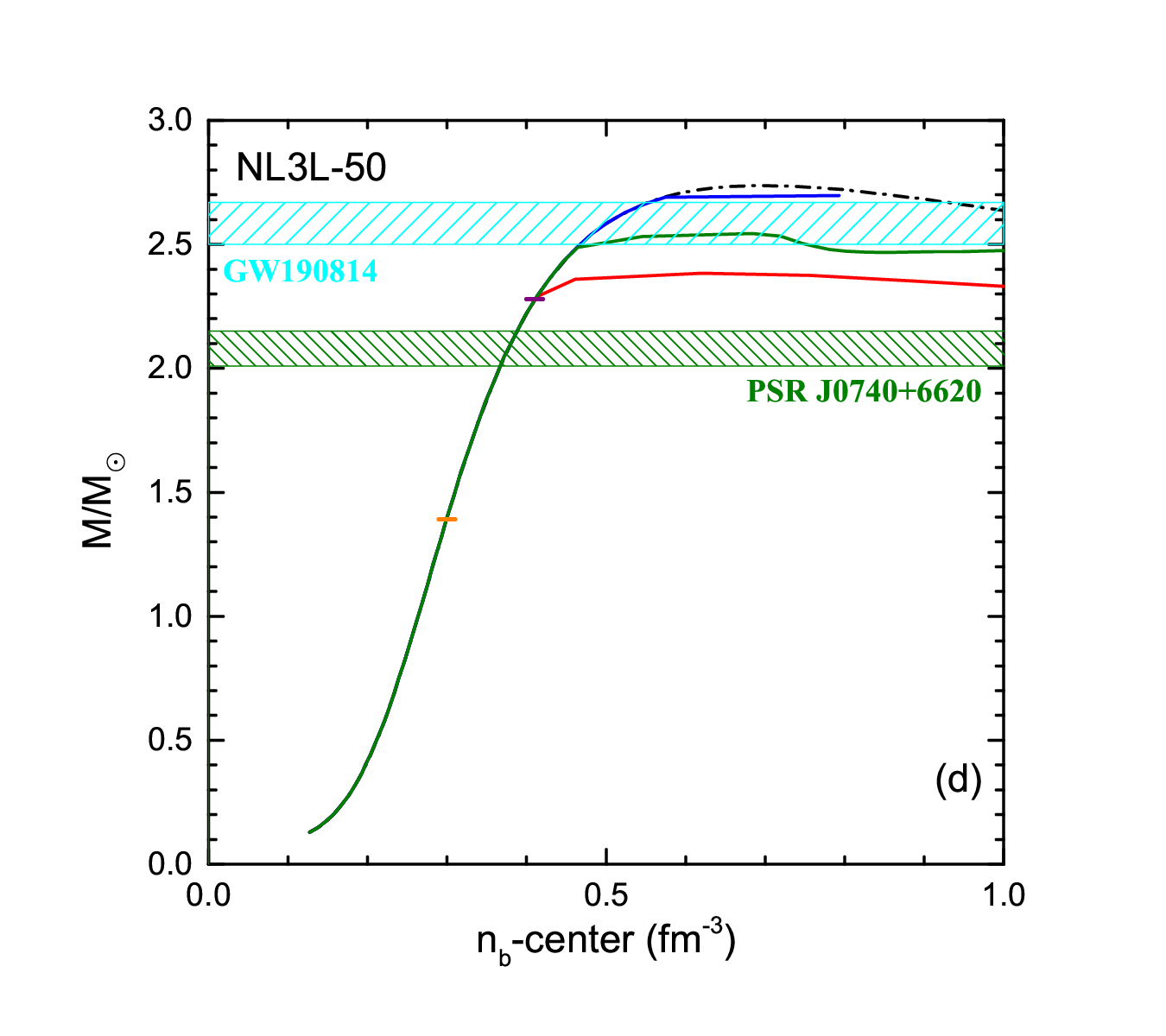}
    \caption{Mass-radius relations and mass-central density relations for different models and couplings.
    The results of pure hadronic EOS (dash-dot lines) are compared with those including H particle for different couplings (same as Fig.~\ref{fig:4-nbp}).
    The shaded areas correspond to simultaneous measurements of the mass and radius range from NICER
for PSR J0030+0451~\cite{Riley2019,Miller2019} and PSR J0740+6620, respectively~\cite{Riley2021,Miller2021}.
The hypothesis that the second component of GW190814 is a NS is also depicted~\cite{Abbott2020}.
The inferred radius constraint from GW170817 is shown by grey line~\cite{Annala2018}.}
    \label{fig:5-mr}
\end{figure*}

\begin{figure*}[hptb]
    \centering
    \includegraphics[viewport=10 10 580 570, width=8 cm,clip]{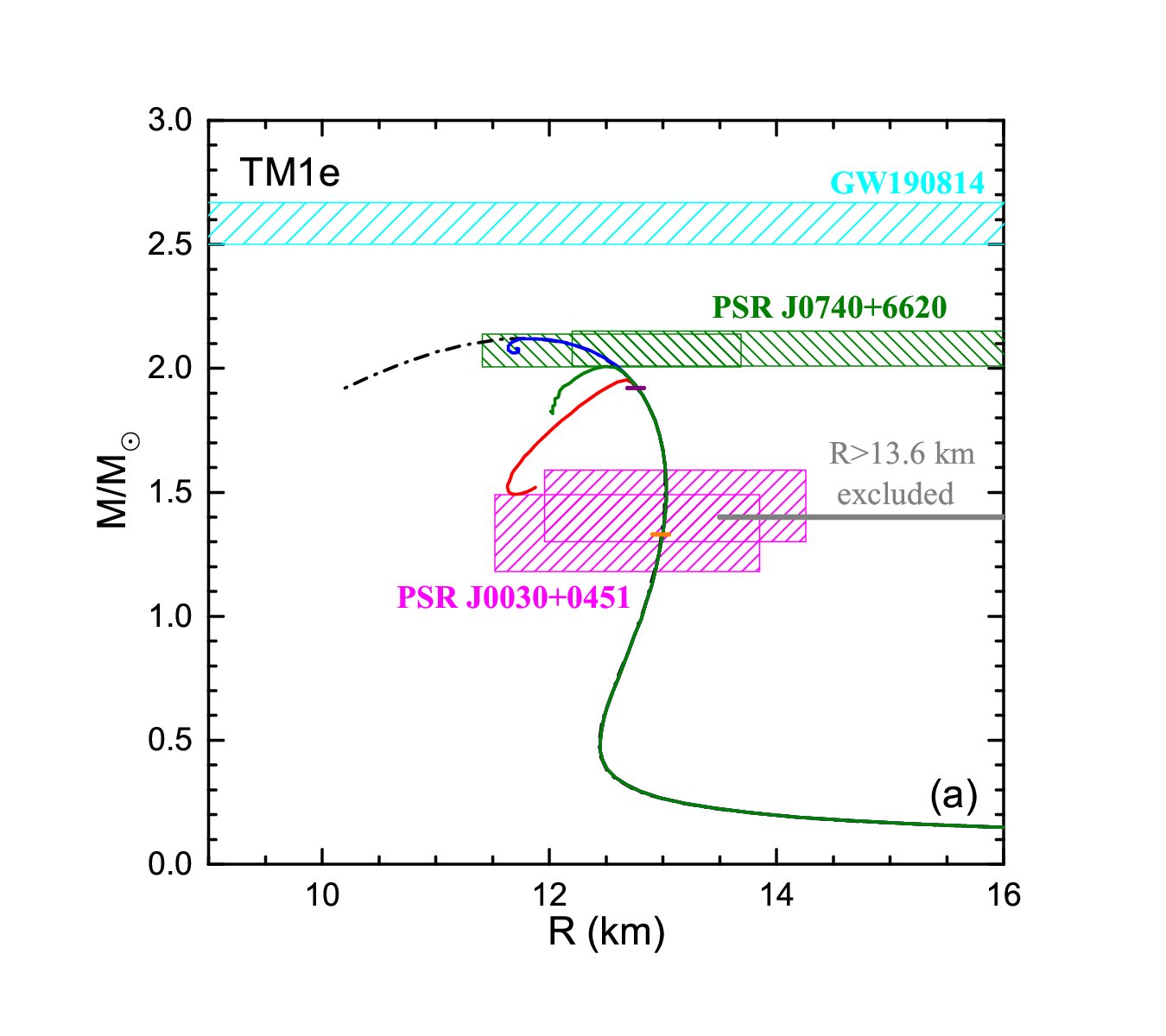}
    \includegraphics[bb=10 10 580 570, width=8 cm,clip]{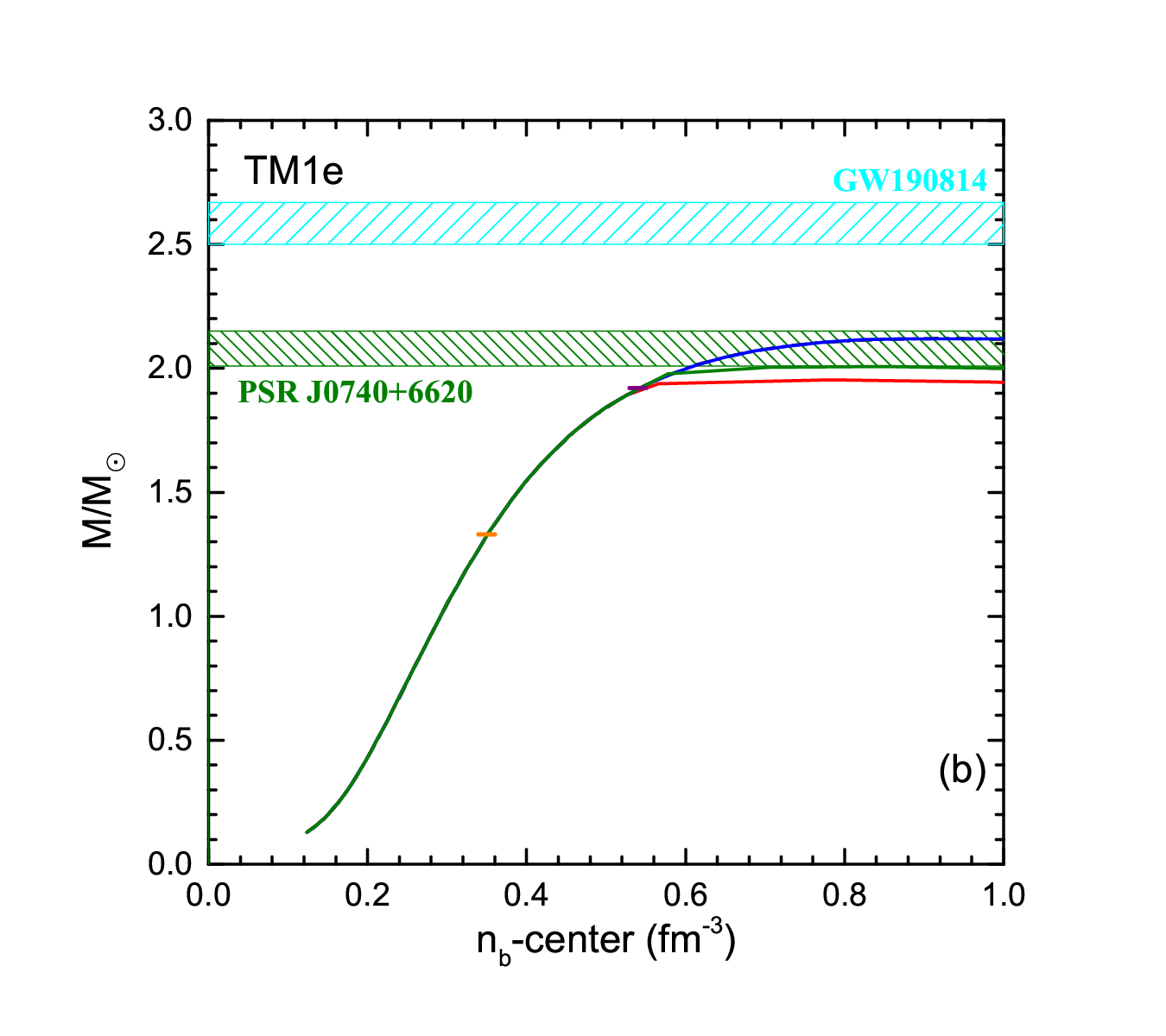} \\
    \includegraphics[viewport=10 10 580 570, width=8 cm,clip]{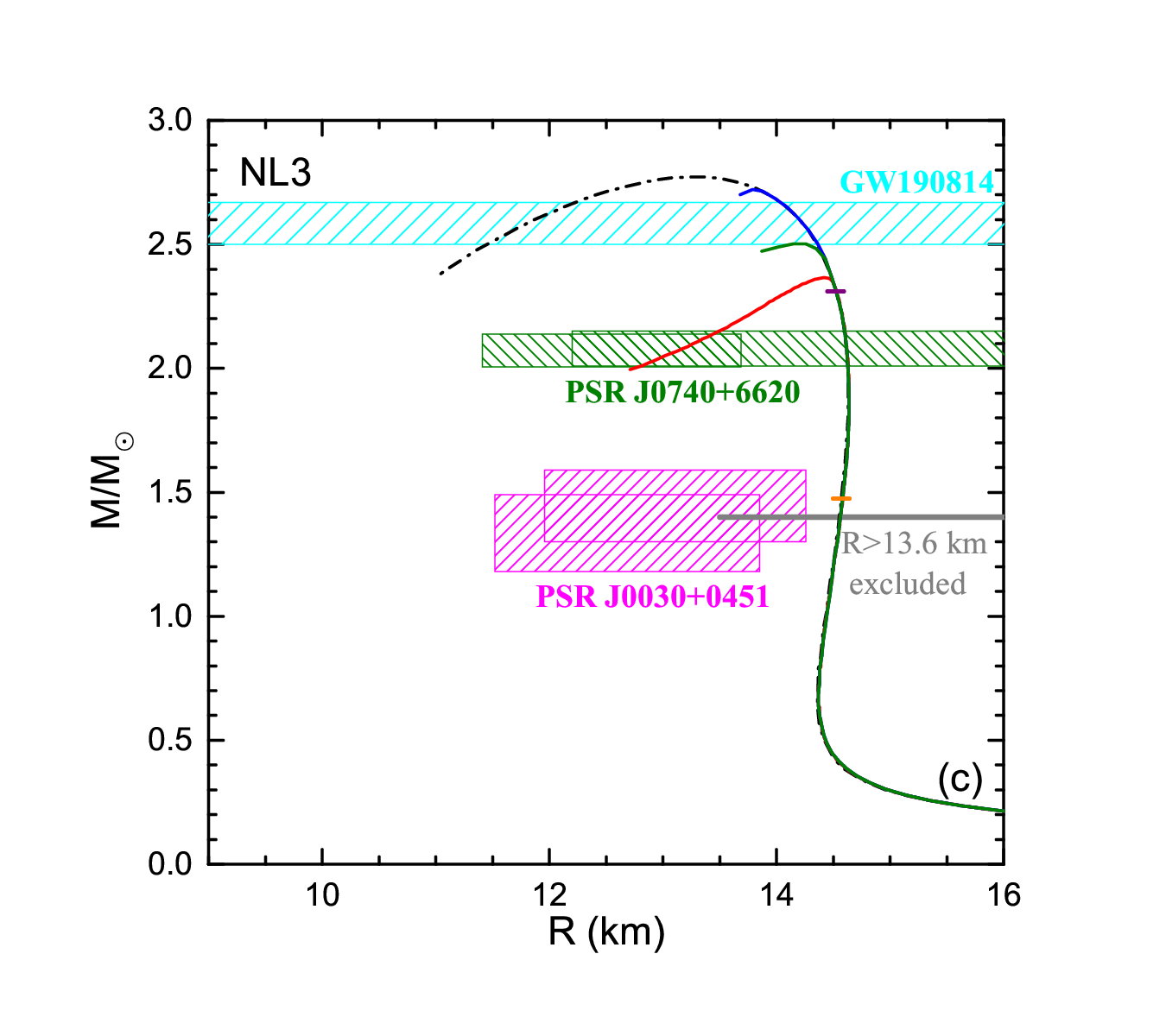}
    \includegraphics[bb=10 10 580 570, width=8 cm,clip]{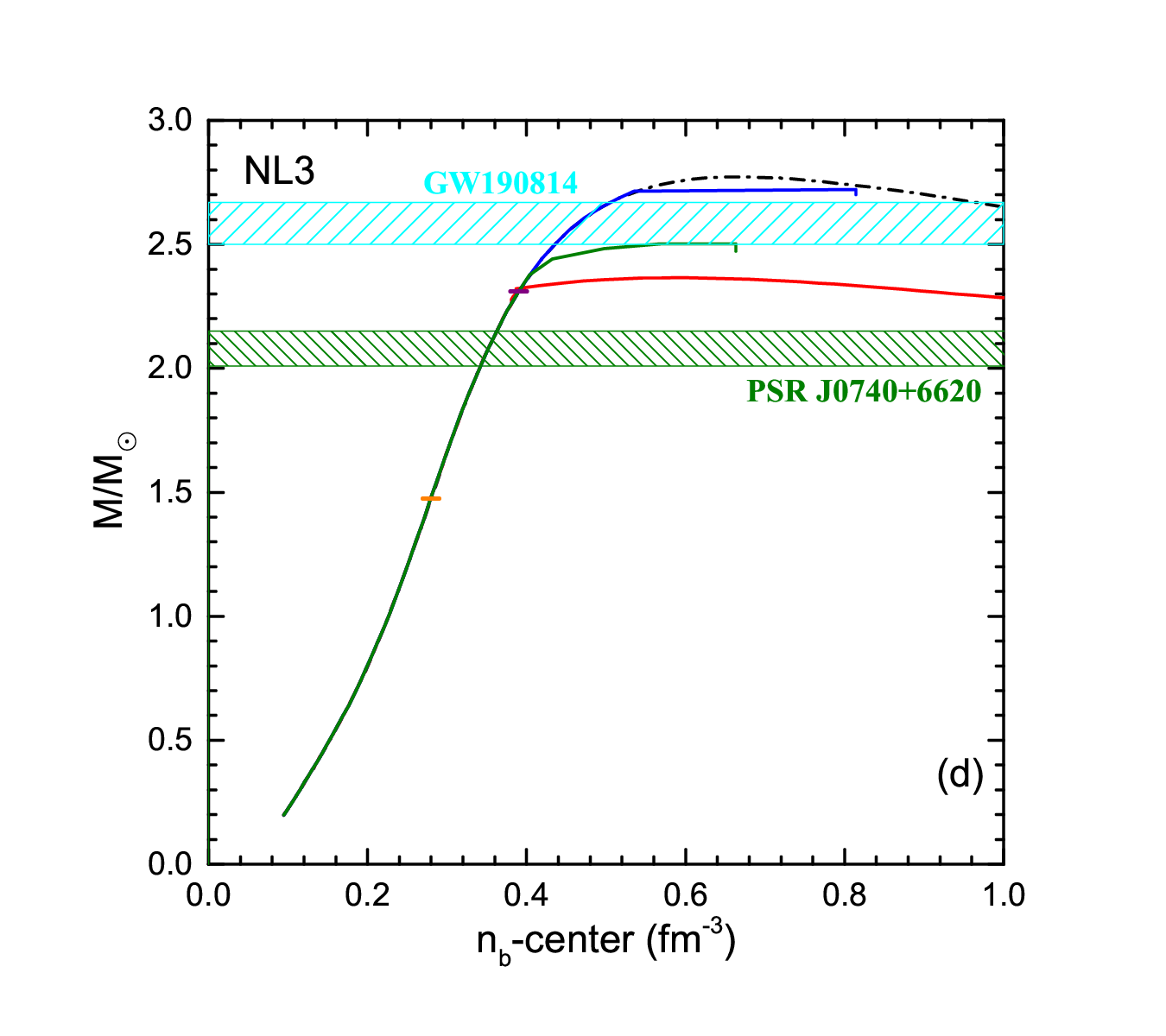}
    \caption{Same as Fig.~\ref{fig:5-mr} but for TM1e and NL3.}
    \label{fig:6-mr}
\end{figure*}

In Figs.~\ref{fig:5-mr} and \ref{fig:6-mr}, we display the mass-radius relations (left panel)
and mass-central density relations (right panel) of NS predicted by using the EOSs utilized in Fig.~\ref{fig:4-nbp}.
Additionally, several constraints derived from astrophysical observations are included in the figures.
With $(\chi_{\sigma H}, \chi_{\omega H})=(0, 0.5)$, the pressure does not decrease soon after the appearance of H particle (i.e. stable, see Fig.~\ref{fig:2-stable}),
but the increase of the pressure does not strong enough to resist gravity and the star would collapse,
which is similar as $(\chi_{\sigma H}, \chi_{\omega H})=(0, 0)$ for NL3L-50.
In the case of $(\chi_{\sigma H}, \chi_{\omega H})=(0, 0)$ for Bigapple, TM1e and NL3, the pressure decreases soon after the appearance of H particle (i.e. unstable, see Fig.~\ref{fig:2-stable}),
so that it leads to the collapse of the star.
Generally the introduction of new degrees of freedom, such as H particle in this work, always reduces the maximum mass of NS.
Consequently, the results of mass and radius exhibit a clear deviation from the pure hadronic mass-radius line.
For $(\chi_{\sigma H}, \chi_{\omega H})=(0.5, 1), (1, 1.5), (0, 1)$,
the results align with the typical addition of new degrees of freedom, like hyperons.
By employing stiff EOSs of BigApple and NL3L-50, these couplings satisfy the constraints from astrophysical observations.
Meanwhile, with TM1e, only $(\chi_{\sigma H}, \chi_{\omega H})=(1, 1.5), (0, 1)$ fulfill the constraints.
Notably, the couplings that adhere to the constraints do not exert an influence on the properties of a $M=1.4~M_\odot$ NS,
because the H particle appears at higher density than its central density.
In the case of no-interaction coupling $(\chi_{\sigma H}, \chi_{\omega H})=(0, 0)$,
the H particle could appear below the central density of $1.4~M_\odot$,
but the star collapses soon thereafter.
The results obtained from NL3 fail to satisfy many of the observed constraints,
as indicated in Fig.~\ref{fig:6-mr},
where NL3 yields excessively large radii due to its large value of symmetry energy
slope $L$.


%
\section{Conclusions}
\label{sec:conclusions}
In this study, we investigated the possibility of H particle's existence in NS.
We first calculated the mass of the H particle using the CMI model,
which predicted a possible $J^P=0^+$ particle with a mass of $2212.7~\rm{MeV}$.
To describe NS matter, we employed the RMF/RMFL model with different parameter sets, namely BigApple, NL3L-50, TM1e, and NL3.
Meanwhile, we explored the presence of the H particle in NS matter considering various choices of attractive and repulsive couplings (${\chi}_{\sigma H}$ and ${\chi}_{\omega H}$).
Our numerical results show some model dependence, but the behavior of H particle shares many commonalities.
Except for the NL3L-50 parameter set, the existence of H particle with purely attractive interaction is not supported.
In these cases, the appearance of H particle leads to the collapse of the NS into a black hole.
Usually a total repulsive interaction is necessary to allow the existence of H particle in the core of NS,
however, a small pure repulsive interaction is not sufficient to support the presence of the H particle in NS by using BigApple, TM1e and NL3.
Interestingly, in the case of NL3L-50, even a slight total attractive interaction of H particle is permissible,
since the pressure supplied by other particles can counterbalance the H particle's attractive potential,
preventing the collapse of the NS.
The presence of H particle generally reduce the maximum mass of NS.
However, with a relatively large total repulsive interaction, the maximum mass can reach 2~$M_\odot$,
allowing for the possibility of H particle to exist in the core of the NS.
On the other hand, a large enough total repulsive interaction can make the system more stable,
but delays the appearance of H particle beyond the center density of NS,
which does not support its existence in NS.

Three types of collapse or instability are addressed in this study.
The first one is the common NS collapse when the mass-radius line exceeds the maximum mass,
where the strong interaction in normal NS matter consisting of $npe\mu$ particles cannot resist the gravity.
Such instability also occur when the H particles are taken into account with (${\chi}_{\sigma H}$, ${\chi}_{\omega H}$)=(0, 1), (1, 1.5) and (0.5, 1).
The second type was discussed in Fig.~\ref{fig:2-stable},
where the presence of H particle causes an immediate decrease in the total pressure,
which induces the stable $npe\mu$ matter to be unstable $npe\mu H$ matter with increasing density.
These choices of (${\chi}_{\sigma H}$, ${\chi}_{\omega H}$) trigger
the collapse after the H particle appears.
The third one is the appearance of H particle does not immediately cause a drop in pressure.
However, a decrease of the derivative of pressure with respect to density means that the interaction among particles becomes
not strong enough to resist gravity and the star collapses.
This was observed in cases (${\chi}_{\sigma H}$, ${\chi}_{\omega H}$)=(0, 0) and (0, 0.5).

In summary, we conclude that from our analysis the presence of H particle within NS is plausible.
To further investigate the exact density at which the H particle appears and its fraction within NS,
one possible approach is to use the quark-meson coupling (QMC) method~\cite{Stone2007,Guichon2018} to calculate the interaction potential,
similar to the reappearance of hyperon potential.
This would provide insights into the behavior of the H particle within NS.
In the present work, we considered H particle as only non-nucleonic degree of freedom.
If other degrees of freedom like hyperons, $\Delta$-isobar or $d^*(2380)$ are taken into account, the results will change.
However, due to the charge neutrality nature of NS, it is expected that these particles would not compete with the H particle in terms of their presence within NS.
Further research is needed to explore these aspects in more detail.

\acknowledgments
This work was supported in part by the National Natural Science Foundation of China under Grants No. 12305148 and No. 12175109,
Hebei Natural Science Foundation No. A2023203055 and No. A202203026,
the Shandong Provincial Natural Science Foundation, China under Grants No. ZR2023QA112,
Fundamental Research Funds for the Central Universities under Grant No. 22CX06006A,
and Postdoctoral Fellowship Program of CPSF under Grant No.GZC20240877, and by Shuimu Tsinghua Scholar Program of Tsinghua University under Grant No.2024SM119.

\appendix
\section{Young-Yamanouchi color $\otimes$ spin basis vectors}\label{sec10}
\setcounter{equation}{0}
\renewcommand\theequation{\Alph{section}.\arabic{equation}}
In this appendix, we present all the possible Young-Yamanouchi color $\otimes$ spin basis vectors in Eqs.(\ref{eq:jp015}-\ref{eq:jp025}).

\begin{widetext}
\begin{align}
J^{P}=0^{+}:&
(\begin{tabular}{|c|c|c|}
\hline
1&2 \\
\cline{1-2}
3&4  \\
\cline{1-2}
\end{tabular}
;
\begin{tabular}{|c|c|c|}
\hline
5  \\
 \cline{1-1}
6 \\
 \cline{1-1}
\end{tabular})_{CS}
=
\sqrt{\frac{1}{2}}
\begin{tabular}{|c|c|c|}
\hline
1&2 \\
\cline{1-2}
3&4  \\
\cline{1-2}
5&6  \\
\cline{1-2}
\end{tabular}_{C_{2}}
\otimes
\begin{tabular}{|c|c|c|}
\hline
1&2&5 \\
\cline{1-3}
3&4&6  \\
\cline{1-3}
\end{tabular}_{S_{18}}
-\sqrt{\frac{1}{2}}
\begin{tabular}{|c|c|c|}
\hline
1&3 \\
\cline{1-2}
2&4  \\
\cline{1-2}
5&6  \\
\cline{1-2}
\end{tabular}_{C_{3}}
\otimes
\begin{tabular}{|c|c|c|}
\hline
1&3&5 \\
\cline{1-3}
2&4&6  \\
\cline{1-3}
\end{tabular}_{S_{20}};\nonumber\\
&
(\begin{tabular}{|c|c|c|}
\hline
1&3 \\
\cline{1-2}
2&4  \\
\cline{1-2}
\end{tabular}
;
\begin{tabular}{|c|c|c|}
\hline
5  \\
 \cline{1-1}
6 \\
 \cline{1-1}
\end{tabular})_{CS}
=
-\sqrt{\frac{1}{2}}
\begin{tabular}{|c|c|c|}
\hline
1&2 \\
\cline{1-2}
3&4  \\
\cline{1-2}
5&6  \\
\cline{1-2}
\end{tabular}_{C_{2}}
\otimes
\begin{tabular}{|c|c|c|}
\hline
1&3&5 \\
\cline{1-3}
2&4&6  \\
\cline{1-3}
\end{tabular}_{S_{20}}
-\sqrt{\frac{1}{2}}
\begin{tabular}{|c|c|c|}
\hline
1&3 \\
\cline{1-2}
2&4  \\
\cline{1-2}
5&6  \\
\cline{1-2}
\end{tabular}_{C_{3}}
\otimes
\begin{tabular}{|c|c|c|}
\hline
1&2&5 \\
\cline{1-3}
3&4&6  \\
\cline{1-3}
\end{tabular}_{S_{18}}.
\label{eq:jp015}
\\
J^{P}=2^{+}:&
(\begin{tabular}{|c|c|c|}
\hline
1&2 \\
\cline{1-2}
3&4  \\
\cline{1-2}
\end{tabular}
;
\begin{tabular}{|c|c|c|}
\hline
5  \\
 \cline{1-1}
6 \\
 \cline{1-1}
\end{tabular})_{CS}
=
\sqrt{\frac{2}{5}}
\begin{tabular}{|c|c|c|}
\hline
1&2 \\
\cline{1-2}
3&4  \\
\cline{1-2}
5&6  \\
\cline{1-2}
\end{tabular}_{C_{2}}
\otimes
\begin{tabular}{|c|c|c|c|c|c|}
\hline
1&2&3&4&6 \\
\cline{1-5}
5  \\
\cline{1-1}
\end{tabular}_{S_{3}}
+\sqrt{\frac{3}{5}}
\begin{tabular}{|c|c|c|}
\hline
1&2 \\
\cline{1-2}
3&4  \\
\cline{1-2}
5&6  \\
\cline{1-2}
\end{tabular}_{C_{2}}
\otimes
\begin{tabular}{|c|c|c|c|c|c|}
\hline
1&2&3&4&5 \\
\cline{1-5}
6  \\
\cline{1-1}
\end{tabular}_{S_{2}};\nonumber\\
&
(\begin{tabular}{|c|c|c|}
\hline
1&3 \\
\cline{1-2}
2&4  \\
\cline{1-2}
\end{tabular}
;
\begin{tabular}{|c|c|c|}
\hline
5  \\
 \cline{1-1}
6 \\
 \cline{1-1}
\end{tabular})_{CS}
=
-\sqrt{\frac{2}{5}}
\begin{tabular}{|c|c|c|}
\hline
1&3 \\
\cline{1-2}
2&4  \\
\cline{1-2}
5&6  \\
\cline{1-2}
\end{tabular}_{C_{3}}
\otimes
\begin{tabular}{|c|c|c|c|c|c|}
\hline
1&2&3&4&6 \\
\cline{1-5}
5  \\
\cline{1-1}
\end{tabular}_{S_{3}}
-\sqrt{\frac{1}{6}}
\begin{tabular}{|c|c|c|}
\hline
1&3 \\
\cline{1-2}
2&5  \\
\cline{1-2}
4&6  \\
\cline{1-2}
\end{tabular}_{C_{4}}
\otimes
\begin{tabular}{|c|c|c|c|c|c|}
\hline
1&3&4&5&6 \\
\cline{1-5}
2  \\
\cline{1-1}
\end{tabular}_{S_{6}}.
\label{eq:jp017}
\\
J^{P}=2^{+}:&
(\begin{tabular}{|c|c|c|}
\hline
1&2 \\
\cline{1-2}
3&4  \\
\cline{1-2}
\end{tabular}
;
\begin{tabular}{|c|c|c|}
\hline
5  \\
 \cline{1-1}
6 \\
 \cline{1-1}
\end{tabular})_{CS}
=
-\sqrt{\frac{1}{3}}
\begin{tabular}{|c|c|c|}
\hline
1&4 \\
\cline{1-2}
2&5  \\
\cline{1-2}
3&6  \\
\cline{1-2}
\end{tabular}_{C_{5}}
\otimes
\begin{tabular}{|c|c|c|c|c|c|}
\hline
1&2&3&5&6 \\
\cline{1-5}
4  \\
\cline{1-1}
\end{tabular}_{S_{6}}
-\sqrt{\frac{1}{6}}
\begin{tabular}{|c|c|c|}
\hline
1&3 \\
\cline{1-2}
2&5  \\
\cline{1-2}
4&6  \\
\cline{1-2}
\end{tabular}_{C_{4}}
\otimes
\begin{tabular}{|c|c|c|c|c|c|}
\hline
1&3&4&5&6 \\
\cline{1-5}
2  \\
\cline{1-1}
\end{tabular}_{S_{6}}
+\sqrt{\frac{1}{6}}
\begin{tabular}{|c|c|c|}
\hline
1&2 \\
\cline{1-2}
3&5  \\
\cline{1-2}
4&6  \\
\cline{1-2}
\end{tabular}_{C_{1}}
\otimes
\begin{tabular}{|c|c|c|c|c|c|}
\hline
1&2&4&5&6 \\
\cline{1-5}
3  \\
\cline{1-1}
\end{tabular}_{S_{5}}\nonumber\\
&\quad \quad \quad \quad \quad \quad
-\sqrt{\frac{1}{3}}
\begin{tabular}{|c|c|c|}
\hline
1&2 \\
\cline{1-2}
3&5  \\
\cline{1-2}
4&6  \\
\cline{1-2}
\end{tabular}_{C_{1}}
\otimes
\begin{tabular}{|c|c|c|c|c|c|}
\hline
1&2&3&5&6 \\
\cline{1-5}
4  \\
\cline{1-1}
\end{tabular}_{S_{4}}
;
\nonumber\\
&
(\begin{tabular}{|c|c|c|}
\hline
1&3 \\
\cline{1-2}
2&4  \\
\cline{1-2}
\end{tabular}
;
\begin{tabular}{|c|c|c|}
\hline
5  \\
 \cline{1-1}
6 \\
 \cline{1-1}
\end{tabular})_{CS}
=
\sqrt{\frac{1}{3}}
\begin{tabular}{|c|c|c|}
\hline
1&4 \\
\cline{1-2}
2&5  \\
\cline{1-2}
3&6  \\
\cline{1-2}
\end{tabular}_{C_{5}}
\otimes
\begin{tabular}{|c|c|c|c|c|c|}
\hline
1&2&4&5&6 \\
\cline{1-5}
3  \\
\cline{1-1}
\end{tabular}_{S_{5}}
-\sqrt{\frac{1}{6}}
\begin{tabular}{|c|c|c|}
\hline
1&3 \\
\cline{1-2}
2&5  \\
\cline{1-2}
4&6  \\
\cline{1-2}
\end{tabular}_{C_{4}}
\otimes
\begin{tabular}{|c|c|c|c|c|c|}
\hline
1&2&4&5&6 \\
\cline{1-5}
3  \\
\cline{1-1}
\end{tabular}_{S_{5}}
-\sqrt{\frac{1}{6}}
\begin{tabular}{|c|c|c|}
\hline
1&2 \\
\cline{1-2}
3&5  \\
\cline{1-2}
4&6  \\
\cline{1-2}
\end{tabular}_{C_{1}}
\otimes
\begin{tabular}{|c|c|c|c|c|c|}
\hline
1&3&4&5&6 \\
\cline{1-5}
2  \\
\cline{1-1}
\end{tabular}_{S_{6}} \nonumber\\
&\quad \quad \quad \quad \quad \quad
-\sqrt{\frac{1}{3}}
\begin{tabular}{|c|c|c|}
\hline
1&3 \\
\cline{1-2}
2&5  \\
\cline{1-2}
4&6  \\
\cline{1-2}
\end{tabular}_{C_{4}}
\otimes
\begin{tabular}{|c|c|c|c|c|c|}
\hline
1&2&3&5&6 \\
\cline{1-5}
4  \\
\cline{1-1}
\end{tabular}_{S_{4}}
\label{eq:jp019}.\\
J^{P}=0^{+}:&
(\begin{tabular}{|c|c|c|}
\hline
1&2 \\
\cline{1-2}
3&4  \\
\cline{1-2}
\end{tabular}
;
\begin{tabular}{|c|c|c|}
\hline
5  \\
 \cline{1-1}
6 \\
 \cline{1-1}
\end{tabular})_{CS}
=
-\sqrt{\frac{1}{3}}
\begin{tabular}{|c|c|c|}
\hline
1&4 \\
\cline{1-2}
2&5  \\
\cline{1-2}
3&6  \\
\cline{1-2}
\end{tabular}_{C_{5}}
\otimes
\begin{tabular}{|c|c|c|c|c|c|}
\hline
1&2&3 \\
\cline{1-3}
4 &5&6 \\
\cline{1-3}
\end{tabular}_{S_{19}}
-\sqrt{\frac{1}{6}}
\begin{tabular}{|c|c|c|}
\hline
1&3 \\
\cline{1-2}
2&5  \\
\cline{1-2}
4&6  \\
\cline{1-2}
\end{tabular}_{C_{4}}
\otimes
\begin{tabular}{|c|c|c|c|c|c|}
\hline
1&3&4 \\
\cline{1-3}
2&5&6  \\
\cline{1-3}
\end{tabular}_{S_{19}}
+\sqrt{\frac{1}{6}}
\begin{tabular}{|c|c|c|}
\hline
1&2 \\
\cline{1-2}
3&5  \\
\cline{1-2}
4&6  \\
\cline{1-2}
\end{tabular}_{C_{1}}
\otimes
\begin{tabular}{|c|c|c|c|c|c|}
\hline
1&2&4 \\
\cline{1-3}
3&5&6  \\
\cline{1-3}
\end{tabular}_{S_{17}}\nonumber\\
&\quad \quad \quad \quad \quad \quad
-\sqrt{\frac{1}{3}}
\begin{tabular}{|c|c|c|}
\hline
1&2 \\
\cline{1-2}
3&5  \\
\cline{1-2}
4&6  \\
\cline{1-2}
\end{tabular}_{C_{1}}
\otimes
\begin{tabular}{|c|c|c|c|c|c|}
\hline
1&2&3 \\
\cline{1-3}
4&5&6  \\
\cline{1-3}
\end{tabular}_{S_{16}}
;\nonumber
\\
&
(\begin{tabular}{|c|c|c|}
\hline
1&3 \\
\cline{1-2}
2&4  \\
\cline{1-2}
\end{tabular}
;
\begin{tabular}{|c|c|c|}
\hline
5  \\
 \cline{1-1}
6 \\
 \cline{1-1}
\end{tabular})_{CS}
=
\sqrt{\frac{1}{3}}
\begin{tabular}{|c|c|c|}
\hline
1&4 \\
\cline{1-2}
2&5  \\
\cline{1-2}
3&6  \\
\cline{1-2}
\end{tabular}_{C_{5}}
\otimes
\begin{tabular}{|c|c|c|c|c|c|}
\hline
1&2&4 \\
\cline{1-3}
3&5&6  \\
\cline{1-3}
\end{tabular}_{S_{17}}
-\sqrt{\frac{1}{6}}
\begin{tabular}{|c|c|c|}
\hline
1&3 \\
\cline{1-2}
2&5  \\
\cline{1-2}
4&6  \\
\cline{1-2}
\end{tabular}_{C_{4}}
\otimes
\begin{tabular}{|c|c|c|c|c|c|}
\hline
1&2&4 \\
\cline{1-3}
3&5&6 \\
\cline{1-3}
\end{tabular}_{S_{17}}
-\sqrt{\frac{1}{6}}
\begin{tabular}{|c|c|c|}
\hline
1&2 \\
\cline{1-2}
3&5  \\
\cline{1-2}
4&6  \\
\cline{1-2}
\end{tabular}_{C_{1}}
\otimes
\begin{tabular}{|c|c|c|c|c|c|}
\hline
1&3&4 \\
\cline{1-3}
2&5&6  \\
\cline{1-3}
\end{tabular}_{S_{19}} \nonumber\\
&\quad \quad \quad \quad \quad \quad
-\sqrt{\frac{1}{3}}
\begin{tabular}{|c|c|c|}
\hline
1&3 \\
\cline{1-2}
2&5  \\
\cline{1-2}
4&6  \\
\cline{1-2}
\end{tabular}_{C_{4}}
\otimes
\begin{tabular}{|c|c|c|c|c|c|}
\hline
1&2&3 \\
\cline{1-3}
4&5&6  \\
\cline{1-3}
\end{tabular}_{S_{16}}.
\label{eq:jp021}
\\
J^{P}=1^{+}:&
(\begin{tabular}{|c|c|c|}
\hline
1&2 \\
\cline{1-2}
3&4  \\
\cline{1-2}
\end{tabular}
;
\begin{tabular}{|c|c|c|}
\hline
5  \\
 \cline{1-1}
6 \\
 \cline{1-1}
\end{tabular})_{CS}
=
\frac{1}{3}
\begin{tabular}{|c|c|c|}
\hline
1&4 \\
\cline{1-2}
2&5  \\
\cline{1-2}
3&6  \\
\cline{1-2}
\end{tabular}_{C_{5}}
\otimes
\begin{tabular}{|c|c|c|c|c|c|}
\hline
1&3&4&5 \\
\cline{1-4}
2&6 \\
\cline{1-2}
\end{tabular}_{S_{13}}
+\frac{\sqrt{2}}{3}
\begin{tabular}{|c|c|c|}
\hline
1&4 \\
\cline{1-2}
2&5  \\
\cline{1-2}
3&6  \\
\cline{1-2}
\end{tabular}_{C_{5}}
\otimes
\begin{tabular}{|c|c|c|c|c|c|}
\hline
1&3&4&6 \\
\cline{1-4}
2&5 \\
\cline{1-2}
\end{tabular}_{S_{14}}
+\sqrt{\frac{1}{18}}
\begin{tabular}{|c|c|c|}
\hline
1&3 \\
\cline{1-2}
2&5  \\
\cline{1-2}
4&6  \\
\cline{1-2}
\end{tabular}_{C_{4}}
\otimes
\begin{tabular}{|c|c|c|c|c|c|}
\hline
1&3&4&5 \\
\cline{1-4}
2&6  \\
\cline{1-2}
\end{tabular}_{S_{13}}
\nonumber\\
&\quad \quad \quad \quad \quad \quad
+\frac{1}{3}
\begin{tabular}{|c|c|c|}
\hline
1&3 \\
\cline{1-2}
2&5  \\
\cline{1-2}
4&6  \\
\cline{1-2}
\end{tabular}_{C_{4}}
\otimes
\begin{tabular}{|c|c|c|c|c|c|}
\hline
1&3&4&6 \\
\cline{1-4}
2&5  \\
\cline{1-2}
\end{tabular}_{S_{14}}
-\sqrt{\frac{1}{18}}
\begin{tabular}{|c|c|c|}
\hline
1&2 \\
\cline{1-2}
3&5  \\
\cline{1-2}
4&6  \\
\cline{1-2}
\end{tabular}_{C_{1}}
\otimes
\begin{tabular}{|c|c|c|c|c|c|}
\hline
1&2&4&5 \\
\cline{1-4}
3&6  \\
\cline{1-2}
\end{tabular}_{S_{10}}
-\frac{1}{3}
\begin{tabular}{|c|c|c|}
\hline
1&2 \\
\cline{1-2}
3&5  \\
\cline{1-2}
4&6  \\
\cline{1-2}
\end{tabular}_{C_{1}}
\otimes
\begin{tabular}{|c|c|c|c|c|c|}
\hline
1&2&4&6 \\
\cline{1-4}
3&5  \\
\cline{1-2}
\end{tabular}_{S_{11}}
\nonumber\\
&\quad \quad \quad \quad \quad \quad
+\sqrt{\frac{1}{3}}
\begin{tabular}{|c|c|c|}
\hline
1&2 \\
\cline{1-2}
3&5  \\
\cline{1-2}
4&6  \\
\cline{1-2}
\end{tabular}_{C_{1}}
\otimes
\begin{tabular}{|c|c|c|c|c|c|}
\hline
1&2&3 &5\\
\cline{1-4}
4&6  \\
\cline{1-2}
\end{tabular}_{S_{8}}
+\frac{\sqrt{2}}{3}
\begin{tabular}{|c|c|c|}
\hline
1&2 \\
\cline{1-2}
3&5  \\
\cline{1-2}
4&6  \\
\cline{1-2}
\end{tabular}_{C_{1}}
\otimes
\begin{tabular}{|c|c|c|c|c|c|}
\hline
1&2&3 &6\\
\cline{1-4}
4&5  \\
\cline{1-2}
\end{tabular}_{S_{9}}
;
\nonumber
\\
&
(\begin{tabular}{|c|c|c|}
\hline
1&3 \\
\cline{1-2}
2&4  \\
\cline{1-2}
\end{tabular}
;
\begin{tabular}{|c|c|c|}
\hline
5  \\
 \cline{1-1}
6 \\
 \cline{1-1}
\end{tabular})_{CS}
=
-\frac{1}{3}
\begin{tabular}{|c|c|c|}
\hline
1&4 \\
\cline{1-2}
2&5  \\
\cline{1-2}
3&6  \\
\cline{1-2}
\end{tabular}_{C_{5}}
\otimes
\begin{tabular}{|c|c|c|c|c|c|}
\hline
1&2&4&5 \\
\cline{1-4}
3&6 \\
\cline{1-2}
\end{tabular}_{S_{10}}
-\frac{\sqrt{2}}{3}
\begin{tabular}{|c|c|c|}
\hline
1&4 \\
\cline{1-2}
2&5  \\
\cline{1-2}
3&6  \\
\cline{1-2}
\end{tabular}_{C_{5}}
\otimes
\begin{tabular}{|c|c|c|c|c|c|}
\hline
1&2&4&6 \\
\cline{1-4}
3&5 \\
\cline{1-2}
\end{tabular}_{S_{11}}
+\sqrt{\frac{1}{18}}
\begin{tabular}{|c|c|c|}
\hline
1&3 \\
\cline{1-2}
2&5  \\
\cline{1-2}
4&6  \\
\cline{1-2}
\end{tabular}_{C_{4}}
\otimes
\begin{tabular}{|c|c|c|c|c|c|}
\hline
1&2&4&5 \\
\cline{1-4}
3&6  \\
\cline{1-2}
\end{tabular}_{S_{10}}
\nonumber\\
&\quad \quad \quad \quad \quad \quad
+\frac{1}{3}
\begin{tabular}{|c|c|c|}
\hline
1&3 \\
\cline{1-2}
2&5  \\
\cline{1-2}
4&6  \\
\cline{1-2}
\end{tabular}_{C_{4}}
\otimes
\begin{tabular}{|c|c|c|c|c|c|}
\hline
1&2&4&6 \\
\cline{1-4}
3&5  \\
\cline{1-2}
\end{tabular}_{S_{11}}
+\sqrt{\frac{1}{18}}
\begin{tabular}{|c|c|c|}
\hline
1&2 \\
\cline{1-2}
3&5  \\
\cline{1-2}
4&6  \\
\cline{1-2}
\end{tabular}_{C_{1}}
\otimes
\begin{tabular}{|c|c|c|c|c|c|}
\hline
1&3&4&5 \\
\cline{1-4}
2&6  \\
\cline{1-2}
\end{tabular}_{S_{13}}
+\frac{1}{3}
\begin{tabular}{|c|c|c|}
\hline
1&2 \\
\cline{1-2}
3&5  \\
\cline{1-2}
4&6  \\
\cline{1-2}
\end{tabular}_{C_{1}}
\otimes
\begin{tabular}{|c|c|c|c|c|c|}
\hline
1&2&4&6 \\
\cline{1-4}
2&6  \\
\cline{1-2}
\end{tabular}_{S_{14}}
\nonumber\\
&\quad \quad \quad \quad \quad \quad
+\frac{1}{3}
\begin{tabular}{|c|c|c|}
\hline
1&3 \\
\cline{1-2}
2&5  \\
\cline{1-2}
4&6  \\
\cline{1-2}
\end{tabular}_{C_{4}}
\otimes
\begin{tabular}{|c|c|c|c|c|c|}
\hline
1&2&3 &5\\
\cline{1-4}
4&6  \\
\cline{1-2}
\end{tabular}_{S_{8}}
+\frac{\sqrt{2}}{3}
\begin{tabular}{|c|c|c|}
\hline
1&3 \\
\cline{1-2}
2&5  \\
\cline{1-2}
4&6  \\
\cline{1-2}
\end{tabular}_{C_{4}}
\otimes
\begin{tabular}{|c|c|c|c|c|c|}
\hline
1&2&3 &6\\
\cline{1-4}
4&5  \\
\cline{1-2}
\end{tabular}_{S_{9}}.
\label{eq:jp025}
\end{align}
\end{widetext}


\newpage
\begin{thebibliography}{}
\bibitem{Jaffe1977}
R. L. Jaffe,
Perhaps a Stable Dihyperon,
Phys. Rev. Lett. 38, \textbf{195}, 617(E) (1977).
\bibitem{Farrar2018}
G. R. Farrar,
A precision test of the nature of Dark Matter and a probe of the QCD phase transition,
arXiv:1805.03723 [hep-ph]
\bibitem{Shahrbaf2022}
M. Shahrbaf, D. Blaschke, S. Typel, G.?R. Farrar, and D.?E. Alvarez-Castillo,
Sexaquark dilemma in neutron stars and its solution by quark deconfinement,
Phys. Rev. D \textbf{105}, 103005 (2022).
\bibitem{Weber2007}
F. Weber, M. Meixner, R. P. Negreiros, and M. Malheiro,
Ultra-Dense Neutron Star Matter, Strange Quark Stars, and the Nuclear Equation of State,
Int. J. Mod. Phys. E \textbf{16}, 1165-1180 (2007).
\bibitem{Stone2007}
J. Rikovska-Stone, P. A. M. Guichon, H. H. Matevosyan, A. W. Thomas,
Cold uniform matter and neutron stars in the quark-mesons-coupling model,
Nucl. Phys. A \textbf{792}, 341-369, (2007).
\bibitem{Panda2004}
P. K. Panda, D. P. Menezes, C. Provid\^encia,
Hybrid stars in the quark-meson coupling model with superconducting quark matter,
Phys. Rev. C \textbf{69}, 025207 (2004).
\bibitem{Katayama2015}
T. Katayama, K. Saito,
Hyperons in neutron stars,
Phys. Lett. B \textbf{747}, 43-47 (2015).
\bibitem{Li2018}
J. J. Li, W. H. Long, and A. Sedrakian,
Implications from GW170817 for $\Delta$-isobar Admixed Hypernuclear Compact Stars,
Eur. Phys. J. A \textbf{54}, 133 (2018).
\bibitem{Huang2022}
K. X. Huang,  J. N. Hu,  Y. Zhang \& H. Shen,
The Hadron-quark Crossover in Neutron Star within Gaussian Process Regression Method,
Nucl. Phys. Rev. \textbf{39}, 2 (2022).
\bibitem{Drago2014}
A. Drago, A. Lavagno, G. Pagliara, D. Pigato,
Early appearance of $\Delta$ isobars in neutron stars,
Phys. Rev. C \textbf{90} 065809 (2014).
\bibitem{Sahoo2018}
H. S. Sahoo, G. Mitra, R. Mishra, P.K. Panda, B.-A. Li,
Neutron star matter with $\Delta$ isobars in a relativistic quark model,
Phys. Rev. C \textbf{98}, 045801 (2018).
\bibitem{Zhu2016}
Z. Y. Zhu, A. Li, J. N. Hu, H. Sagawa,
$\Delta$ (1232) effects in density-dependent relativistic Hartree-Fock theory and neutron stars
Phys. Rev. C \textbf{94} 045803 (2016).
\bibitem{Li2020}
J. J. Li, A. Sedrakian, F. Weber,
Rapidly rotating $\Delta$-resonance-admixed hypernuclear compact stars,
Phys. Lett. B \textbf{810}, 135812  (2020).
\bibitem{Raduta2021}
A. R. Raduta,
$\Delta$-admixed neutron stars: Spinodal instabilities and dUrca processes,
Phys. Lett. B \textbf{814} 136070 (2021).
\bibitem{Mantziris2020}
A. Mantziris, A. Pastore, I. Vida\~{n}a, D. P. Watts, M. Bashkanov, and A. M. Romero,
Neutron star matter equation of state including $d^*$-hexaquark degrees of freedom,
A\&A \textbf{638}, A40 (2020).
\bibitem{Faessler1997}
A. Faessler, A. J. Buchmann, and M. I. Krivoruchenko,
Nuclear Matter with a Bose Condensate of Dibaryons in Relativistic Mean-Field Theory,
Phys. Lett. B \textbf{391}, 255-260 (1997).
\bibitem{Faessler1997b}
A. Faessler, A. J. Buchmann, and M. I. Krivoruchenko,
Constraints on the $\omega$- and $\sigma$-meson coupling constants with dibaryons,
Phys. Rev. C  \textbf{56}, 1576 (1997).
\bibitem{Faessler1998}
A. Faessler, A. Buchmann, M. Krivoruchenko, and B. Martemyanov,
Dibaryons in nuclear matter,
J. Phys. G, \textbf{24}, 791 (1998).
\bibitem{Glendenning1998}
N. K. Glendenning and J. Schaffner-Bielich,
Neutron star constraints on the H dibaryon,
Phys. Rev. C \textbf{58}, 1298 (1998).
\bibitem{Glendenning2001} 
N. K. Glendenning,
Phase transitions and crystalline structures in neutron star cores,
Phys. Rep. \textbf{342}, 393 (2001).
\bibitem{Heiselberg2000} 
H. Heiselberg and M. Hjorth-Jensen,
Phases of dense matter in neutron stars,
Phys. Rep. \textbf{328}, 237 (2000).
\bibitem{Weber2005} 
F. Weber,
Strange Quark Matter and Compact Stars,
Prog. Part. Nucl. Phys. \textbf{54}, 193 (2005).
\bibitem{Benic2014}
S. Benic, D. Blaschke, D. E. Alvarez-Castillo, T. Fischer and S. Typel,
A new quark-hadron hybrid equation of state for astrophysics,
A\&A  {\bf 577}, A40 (2015)
\bibitem{Nakazato2008} 
K. Nakazato, K. Sumiyoshi, and S. Yamada,
Astrophysical implications of equation of state for hadron-quark mixed phase: Compact stars and stellar collapses,
Phys. Rev. D \textbf{77}, 103006 (2008).
\bibitem{Wu2019}
X. H. Wu and H. Shen,
Nuclear symmetry energy and hadron-quark mixed phase in neutron stars,
Phys. Rev. C \textbf{99}, 065802 (2019).
\bibitem{Ju2021}
M. Ju, J. N. Hu, H. Shen,
Hadron-quark Pasta Phase in Massive Neutron Stars,
Astrophys. J. \textbf{923}(2), 250 (2021).
\bibitem{Vidana2018}
I. Vida\~{n}a, M. Bashkanov, D.P. Watts, A. Pastore,
The $d^*$(2380) in Neutron Stars ? A New Degree of Freedom?
Phys. Lett. B \textbf{781}, 112-116 (2018).
\bibitem{Bashkanov2019}
M. Bashkanov, S. Kay, D. Watts, et al.,
Deuteron photodisintegration by polarized photons in the region of the $d^*$(2380),
Phys. Lett. B \textbf{789}, 7-12 (2019).
\bibitem{Kim2020}
H. Kim, K. S. Kim, and M. Oka,
Hexaquark picture for $d^*$(2380),
Phys. Rev. D \textbf{102}, 074023 (2020).
\bibitem{Celi2024}
M. O. Celi, M. Bashkanov, M. Mariani, M. G. Orsaria, A. Pastore, I. F. Ranea-Sandoval, and F. Weber,
Destabilization of high-mass neutron stars by the emergence of $d^*$-hexaquarks,
Phys. Rev. D \textbf{109}, 023004 (2024).
\bibitem{Adlarson2014}
P. Adlarson, M. Bashkanov, H. Clement, et al.,
Evidence for a New Resonance from Polarized Neutron-Proton Scattering,
Phys. Rev. Lett. \textbf{112}, 202301 (2014).
\bibitem{Witten1984}
E. Witten,
Cosmic separation of phases,
Phys. Rev. D \textbf{30}, 272 (1984).
\bibitem{Liu2023}
H. Liu, Y.-H. Yang, Y. Han, and P.-C. Chu,
Properties of quark matter and hybrid stars from a quasiparticle model,
Phys. Rev. D \textbf{108}, 034004 (2023).
\bibitem{Fonseca2021}
E. Fonseca, H.T. Cromartie, T.T. Pennucci, et al.,
Refined Mass and Geometric Measurements of the High-mass PSR J0740+6620
Astrophys. J. Lett. \textbf{915}, L12 (2021).
%
\bibitem{Cromartie2020}
H. T. Cromartie, E. Fonseca, S. M. Ransom, et al.,
Relativistic Shapiro delay measurements of an extremely massive millisecond pulsar,
Nat. Astron. \textbf{4}, 72 (2020).
\bibitem{Romani2022}
R. W. Romani, D. Kandel, A.V. Filippenko, T.G. Brink, W. Zheng,
PSR J0952?0607: The Fastest and Heaviest Known Galactic Neutron Star,
Astrophys. J. Lett. \textbf{934}, L18 (2022).
%
\bibitem{Arzoumanian2018}
Z. Arzoumanian et al.,
The NANOGrav 11-year Data Set: High-precision Timing of 45 Millisecond Pulsars,
Astrophys. J. Suppl. \textbf{235}, 37 (2018).
%
\bibitem{Antoniadis2013}
J. Antoniadis, P. C. C. Freire, N. Wex, et al.,
A Massive Pulsar in a Compact Relativistic Binary,
Sci. \textbf{340}, 448 (2013).
\bibitem{Riley2019}
T. E. Riley et al.,
A NICER View of PSR J0030+0451: Millisecond Pulsar Parameter Estimation,
Astrophys. J. Lett. \textbf{887}, L21 (2019).
\bibitem{Miller2019}
M. C. Miller et al.,
PSR J0030+0451 Mass and Radius from NICER Data and Implications for the Properties of Neutron Star Matter,
Astrophys. J. Lett. \textbf{887}, L24 (2019).
\bibitem{Riley2021}
T. E. Riley, A. L. Watts, P. S. Ray et al.,
A NICER View of the Massive Pulsar PSR J0740+6620 Informed by Radio Timing and XMM-Newton Spectroscopy,
Astrophys. J. Lett. \textbf{918}, L27 (2021).
\bibitem{Miller2021}
M. C. Miller, F. K. Lamb, A. J. Dittmann et al.,
The Radius of PSR J0740+6620 from NICER and XMM-Newton Data,
Astrophys. J. Lett. \textbf{918}, L28 (2021).
%
\bibitem{Abbott2017}
B. P. Abbott, et al.(The LIGO Scientific Collaboration and the Virgo Collaboration),
GW170817: Observation of Gravitational Waves from a Binary Neutron Star Inspiral,
Phys. Rev. Lett. \textbf{119}, 161101 (2017).
%
\bibitem{Annala2018}
E. Annala, T. Gorda, A. Kurkela, A. Vuorinen,
Gravitational-Wave Constraints on the Neutron-Star-Matter Equation of State,
Phys. Rev. Lett. \textbf{120}, 172703 (2018).
\bibitem{Abbott2020}
R. Abbott et al. (LIGO Scientific and Virgo Collaborations),
GW190814: Gravitational Waves from the Coalescence of a 23 Solar Mass Black Hole with a 2.6 Solar Mass Compact Object,
Astrophys. J. Lett. \textbf{896}, L44 (2020).
\bibitem{Spinella2017}
W. M. Spinella, 
A Systematic Investigation of Exotic Matter in Neutron Stars, 
Ph.D. thesis, Claremont Graduate University \& San Diego State University (2017).
\bibitem{Sedrakian2023}
A. Sedrakian, J. J. Li, F. Weber,
Heavy Baryons in Compact Stars
Prog. Part. Nucl. Phys. \textbf{131}, 104041  (2023).
\bibitem{Liu:2019zoy}
Y.~R.~Liu, H.~X.~Chen, W.~Chen, X.~Liu and S.~L.~Zhu,
Pentaquark and Tetraquark states,
Prog. Part. Nucl. Phys. \textbf{107}, 237-320 (2019).
\bibitem{Weng:2019ynv}
X.~Z.~Weng, X.~L.~Chen, W.~Z.~Deng and S.~L.~Zhu,
Hidden-charm pentaquarks and $P_c$ states,
Phys. Rev. D \textbf{100} no.1, 016014 (2019).
%
\bibitem{An:2021vwi}
H.~T.~An, K.~Chen, Z.~W.~Liu and X.~Liu,
Heavy flavor pentaquarks with four heavy quarks,
Phys. Rev. D \textbf{103} no.11, 114027 (2021).
%
\bibitem{Weng:2020jao}
X.~Z.~Weng, X.~L.~Chen, W.~Z.~Deng and S.~L.~Zhu,
Systematics of fully heavy tetraquarks,
Phys. Rev. D \textbf{103} (2021) no.3, 034001.
%
\bibitem{Weng:2021ngd}
X.~Z.~Weng, W.~Z.~Deng and S.~L.~Zhu,
Triply heavy tetraquark states,
Phys. Rev. D \textbf{105} (2022) no.3, 034026.
%
\bibitem{Liu:2023rzu}
Z.~Liu, H.~T.~An, Z.~W.~Liu and X.~Liu,
Doubly charmed dibaryon states*,
Chin. Phys. C \textbf{47} (2023) no.6, 063105.
%
\bibitem{Karliner:2014gca}
M.~Karliner and J.~L.~Rosner,
Baryons with two heavy quarks: Masses, production, decays, and detection,
Phys. Rev. D \textbf{90} (2014) no.9, 094007.
%
\bibitem{LHCb:2017iph}
R.~Aaij \textit{et al.} [LHCb],
Observation of the doubly charmed baryon $\Xi_{cc}^{++}$,
Phys. Rev. Lett. \textbf{119} (2017) no.11, 112001.


\bibitem{Karliner:2017qjm}
M.~Karliner and J.~L.~Rosner,
Discovery of doubly-charmed $\Xi_{cc}$ baryon implies a stable ($b b \bar{u} \bar{d}$) tetraquark,
Phys. Rev. Lett. \textbf{119} (2017) no.20, 202001.

\bibitem{Eichten:2017ffp}
E.~J.~Eichten and C.~Quigg,
Heavy-quark symmetry implies stable heavy tetraquark mesons $Q_iQ_j \bar q_k \bar q_l$,
Phys. Rev. Lett. \textbf{119} (2017) no.20, 202002.

\bibitem{Luo:2017eub}
S.~Q.~Luo, K.~Chen, X.~Liu, Y.~R.~Liu and S.~L.~Zhu,
Exotic tetraquark states with the $qq\bar{Q}\bar{Q}$ configuration,
Eur. Phys. J. C \textbf{77} (2017) no.10, 709.
%
\bibitem{Weng:2018mmf}
X.~Z.~Weng, X.~L.~Chen and W.~Z.~Deng,
Masses of doubly heavy-quark baryons in an extended chromomagnetic model,
Phys. Rev. D \textbf{97} (2018) no.5, 054008.


\bibitem{Weng:2021hje}
X.~Z.~Weng, W.~Z.~Deng and S.~L.~Zhu,
Doubly heavy tetraquarks in an extended chromomagnetic model *,
Chin. Phys. C \textbf{46} (2022) no.1, 013102.

\bibitem{An:2020vku}
H.~T.~An, K.~Chen and X.~Liu,
Manifestly exotic pentaquarks with a single heavy quark,
Phys. Rev. D \textbf{105} (2022) no.3, 034018.

\bibitem{Liu:2021gva}
Z.~Liu, H.~T.~An, Z.~W.~Liu and X.~Liu,
Where are the hidden-charm hexaquarks?,
Phys. Rev. D \textbf{105} (2022) no.3, 034006.
\bibitem{Park:2017jbn}
W.~Park, A.~Park, S.~Cho and S.~H.~Lee,
$P_c(4380)$ in a constituent quark model,
Phys. Rev. D \textbf{95} no.5, 054027 (2017).
\bibitem{Park:2015nha}
W.~Park, A.~Park and S.~H.~Lee,
Dibaryons in a constituent quark model,
Phys. Rev. D \textbf{92} no.1, 014037 (2015).
\bibitem{An:2020jix}
H.~T.~An, K.~Chen, Z.~W.~Liu and X.~Liu,
Fully heavy pentaquarks,
Phys. Rev. D \textbf{103} no.7, 074006 (2021).
\bibitem{Park:2016mez}
A.~Park, W.~Park and S.~H.~Lee,
Dibaryons with two strange quarks and one heavy flavor in a constituent quark model,
Phys. Rev. D \textbf{94} (2016) no.5, 054027.
\bibitem{Park:2016cmg}
W.~Park, A.~Park and S.~H.~Lee,
Dibaryons with two strange quarks and total spin zero in a constituent quark model,
Phys. Rev. D \textbf{93} (2016) no.7, 074007.

\bibitem{Stancu:1999qr}
F.~Stancu and S.~Pepin,
Isoscalar factors of the permutation group,
Few Body Syst. \textbf{26}, 113-133 (1999).

%
\bibitem{Fattoyev2020} 
F. J. Fattoyev, C. J. Horowitz, J. Piekarewicz, and B. Reed,
GW190814: Impact of a 2.6 solar mass neutron star on the nucleonic equations of state,
Phys. Rev. C \textbf{102}, 065805 (2020).
\bibitem{Wu2021} 
X. H. Wu, S. S. Bao, H. Shen, R. X. Xu,
Effect of the symmetry energy on the secondary component of GW190814 as a neutron star,
Phys. Rev. C \textbf{104}, 015802 (2021).
\bibitem{Ju2025} 
M. Ju, X. H. Wu, H. Shen,
Effects of symmetry energy on the properties of a hadron-quark mixed phase in hybrid stars,
Phys. Rev. C \textbf{111}, 055801 (2025).
\bibitem{Bao2014} 
S. S. Bao, J. N. Hu, Z. W. Zhang, and H. Shen,
Effects of the symmetry energy on properties of neutron star crusts near the neutron drip density,
Phys. Rev. C \textbf{90}, 045802 (2014).
\bibitem{Wu2018} X. H. Wu, A. Ohnishi, and H. Shen,
Effects of quark-matter symmetry energy on hadron-quark coexistence in neutron-star matter,
Phys. Rev. C \textbf{98}, 065801 (2018).
\bibitem{Lalazissis1997} G. A. Lalazissis, J. K\"{o}nig and P. Ring,
New parametrization for the Lagrangian density of relativistic mean field theory,
Phys. Rev. C \textbf{55}, 540 (1997).
%
\bibitem{Guichon2018} P. A. M. Guichon, J. R. Stone, A. W. Thomas,
Quark-Meson-Coupling (QMC) model for finite nuclei, nuclear matter and beyond,
Prog. Part. Nucl. Phys. \textbf{100}, 262 (2018).


\end{thebibliography}
\end{document}